\documentclass[]{aastex63}

\received{December 7, 2020}
\revised{August 20, 2021}
\accepted{September 1, 2021}
\submitjournal{ApJ}

\shorttitle{Tracking Type II Bursts}
\shortauthors{Hegedus et al.}

\usepackage{url} 

\graphicspath{{./}{figures/}}

\begin{document}

\title{Tracking the Source of Solar Type II Bursts through Comparisons of Simulations and Radio Data}

\correspondingauthor{Alexander M. Hegedus}
\email{alexhege@umich.edu}

\author[0000-0001-6247-6934]{Alexander M. Hegedus}
\affiliation{University of Michigan, Department of Climate and Space Sciences and Engineering, Ann Arbor, Michigan, USA}

\author[0000-0003-0472-9408]{Ward B. Manchester IV}
\affiliation{University of Michigan, Department of Climate and Space Sciences and Engineering, Ann Arbor, Michigan, USA}

\author[0000-0002-7077-930X]{Justin C. Kasper}
\affiliation{University of Michigan, Department of Climate and Space Sciences and Engineering, Ann Arbor, Michigan, USA}












\begin{abstract}

The most intense solar energetic particle events are produced by coronal mass ejections (CMEs) accompanied by intense type II radio bursts below 15 MHz.  Understanding where these type II bursts are generated relative to an erupting CME would reveal important details of particle acceleration near the Sun, but the emission cannot be imaged on Earth due to distortion from its ionosphere.  Here, a technique is introduced to identify the likely source location of the emission by comparing the observed dynamic spectrum observed from a single spacecraft against synthetic spectra made from hypothesized emitting regions within a magnetohydrodynamic (MHD) numerical simulation of the recreated CME.  The radio-loud 2005 May 13 CME was chosen as a test case, with Wind/WAVES radio data used to frame the inverse problem of finding the most likely progression of burst locations.  An MHD recreation is used to create synthetic spectra for various hypothesized burst locations.  A framework is developed to score these synthetic spectra by their similarity to the type II frequency profile derived from Wind/WAVES data.  Simulated areas with 4x enhanced entropy and elevated de Hoffmann Teller velocities are found to produce synthetic spectra similar to spacecraft observations.  A geometrical analysis suggests the eastern edge of the entropy derived shock around $(-30^{\circ}, 0^{\circ})$ was emitting in the first hour of the event before falling off, and the western/southwestern edge of the shock centered around $(6^{\circ}, -12^{\circ})$ was a dominant area of radio emission for the 2 hours of simulation data out to 20 solar radii.

\end{abstract}



\section{Introduction} \label{sec:intro}

\begin{figure*}[b]
\centering
\includegraphics[width=0.8\textwidth]{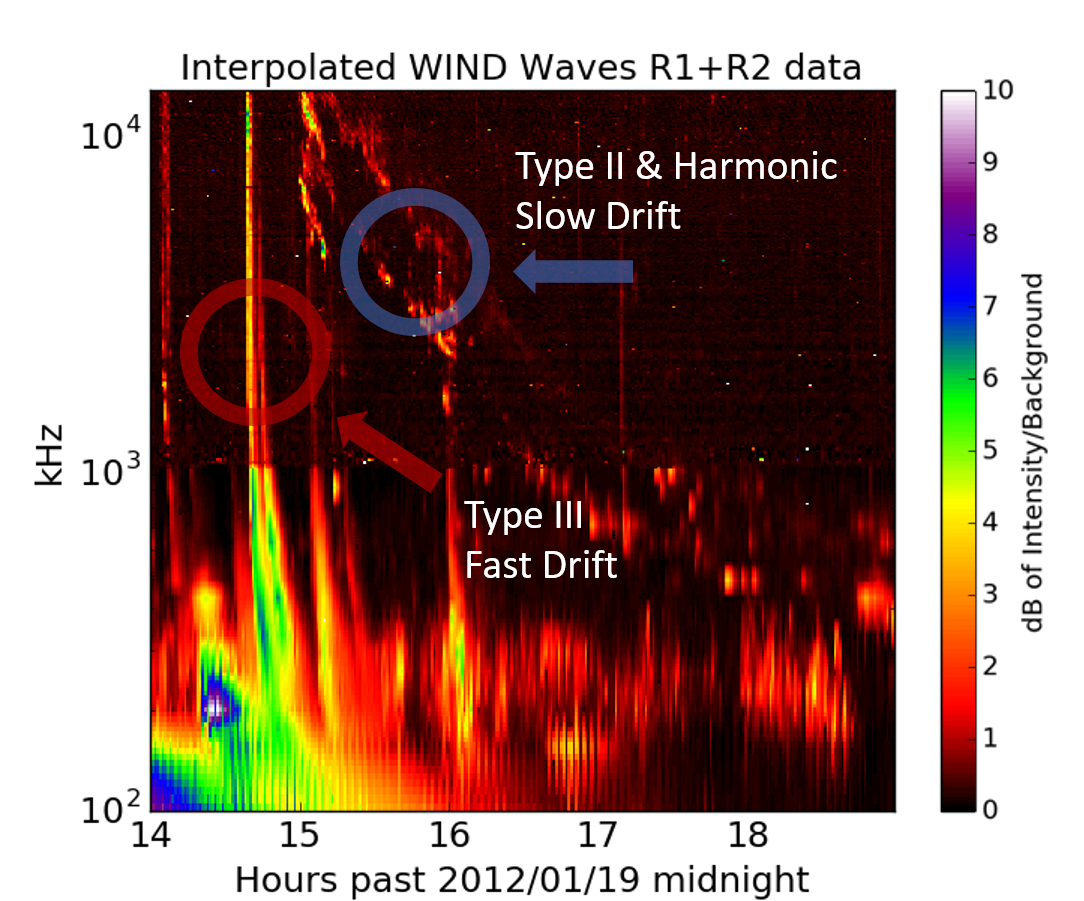}
\caption{\textmd{Radio Spectral Data from Wind/WAVES.  Type II and III bursts are identified by their \replaced{rate of descent in frequency.  Type II bursts follow strong CMEs traveling out from the Sun, and descend relatively slowly, while type III bursts follow a relativistic electron jet that descends far more quickly due to the higher speed of the jet.}{narrow-band structure that drifts to lower frequencies, with type II bursts drifting more slowly than type IIIs.  }}}
\label{fig_wavesii}
\end{figure*}

It has long been known that the solar corona is capable of producing a variety of intense radio bursts in narrow frequency bands with different characteristic variations in frequency, time, and intensity depending on the source \citep{Wild1963}.  Among these bursts are type II and type III radio bursts, which were first observed in the 1950s with ground based radio observatories in frequencies slightly above the 15 MHz ionospheric cutoff \citep{Wild1950}.  These observations continued into the space age with single spacecraft observations like Wind \citep{Bougeret1995} that observed these bursts in the lowest frequencies below 15 MHz.  A Wind \replaced{spectra}{spectrum} in Figure \ref{fig_wavesii} illustrates typical examples of both type II and III bursts, identified by their \replaced{descending narrow-band structure, with type II bursts descending more slowly than type IIIs.}{narrow-band structure that drifts to lower frequencies, with type II bursts drifting more slowly than type IIIs.}  Type II and III solar radio bursts have been observed to be associated with the main types of Solar energetic particles (SEP) events: gradual and impulsive respectively (see review by \cite{Reames2013}).  Type III bursts are more numerous than type II bursts \added{\citep{Cane2002}}, \replaced{but are associated with a minority of the total SEP flux.  Conversely, type II bursts are rarer than type III bursts, but are associated with the bulk of the total SEP flux seen at Earth.}{ but last a shorter amount of time and have a lower SEP fluence than the gradual event associated type II bursts \citep{Reames1999}.}  


\replaced{Solar energetic particles}{SEPs} consist of high-energy ions and electrons, generally in the keV to GeV range, and are thought to be produced by solar flares and coronal mass ejections (CMEs) \citep{Tsurutani2009}. The most intense SEP events are caused by CMEs as they accelerate away from the Sun and plow into the extended solar corona \citep{Reames2013}. These \deleted{CME-associated }SEPs are a form of space weather that poses a threat to both satellites and humans in space.  Energetic particles can access the Earth system through the polar regions, potentially causing harm to satellites and passengers on jets in polar routes.  The higher the flux of incoming energetic particles, the further down in latitude their effects can spread.  In near-Earth space, energetic particles can cause satellite damage and harm astronauts that are not protected by Earth's atmosphere or other means \citep{SSWE2008}.  Interplanetary CMEs (ICMEs) and their shock and sheath regions are also the leading drivers of geomagnetic storms \citep{Ontiveros2010, Kilpua2019}.  Understanding the physics behind these events is \replaced{vital}{crucial} to predicting their impact ahead of time to protect human lives and property, and untangling the relationship between SEPs and type II and III radio bursts is instrumental in this endeavor.   

Electron beams associated with type III bursts were tracked in the 1970s by groups such as \cite{Frank1972} and \cite{Alvarez1972} who found Langmuir \added{waves} associated with the electron beams were causing the type III emission \citep{Gurnett1976, Gurnett1977}.  In addition, \cite{Lin1981, Lin1986} showed that these plasma oscillations occurred alongside a bump-on-tail electron energy distribution, indicating that a plasma emission mechanism was behind the type III radio emission.  This same mechanism is thought to be behind the emission of type II bursts as well \citep{Thejappa_2007}.  

The prevailing physical theory for the plasma emission mechanism was first put forward in 1958 by \cite{Ginzburg58}, which has since been refined by them and many others over the years, as reviewed by \cite{melrose_2008} and references therein.  The basic idea is that the radio bursts are created in a multi-step process that starts with an electron beam, or a sub-population of electrons that are much faster than their neighbors.  \replaced{In the case of type III associated impulsive SEP events, relatively small packets of jettisoned particles from flaring active regions on the surface of the Sun make up the beam with typical (ion) energies of 10 keV to low GeV \citep{Tsurutani2009}.  For type II associated gradual SEP events, the beam is thought to be formed by particles somehow energized by a CME \citep{Cane1984, Cane1987} with typical (ion) energies of 10 keV to 10 MeV \citep{Tsurutani2009}.}{In the case of type III bursts, relatively small packets of jettisoned particles from flaring active regions on the surface of the Sun make up the beam with typical electron energies of 10 - 100 keV \citep{Reames2013}.  For type II bursts, the beam is thought to be formed by particles somehow energized by a CME/interplanetary shock \citep{Cane1984, Cane1987} with typical electron energies on the order of $\sim 1$ MeV \citep{Cliver_2007}.} Once the electron beam forms, it is unstable to wave-particle interactions that can generate Langmuir waves with frequencies around the local plasma frequency via the standard “bump-on-tail” instability.  The waves that are the right frequency to resonate with the beamed electrons convert energy from the electron beam to the plasma waves, relaxing the distribution function to a smoother profile.   Through a combination of three wave decay, ion scattering, and coalescence, a fraction of the power in the Langmuir waves can be converted into radio waves at the corresponding local plasma frequency, or its harmonic.  These radio waves may then escape from their source region and propagate into space. 

The local plasma frequency is only dependent on the density of electrons $n_e$, being proportional to $\sqrt{n_e}$, and so the frequency of the burst trails down through frequency space as the source energetic electrons soar outwards from the Sun.  An example of a Wind/WAVES radio \replaced{spectral}{spectrum} \citep{Bougeret1995} \replaced{showing}{containing} both types of emission is shown in Figure \ref{fig_wavesii}.  Both types of bursts are identified by their \replaced{descending narrow-band structure, and may be differentiated by their rate of descent in frequency, with type II bursts descending more slowly than type IIIs.}{narrow-band structure that drifts to lower frequencies, and may be differentiated by their frequency drift rate, with type II bursts drifting more slowly than type IIIs.}  The full formula for the local plasma frequency $\omega_{pe}$ is shown in Equation \ref{eqn_plasfreq}, where the elementary charge of an electron $e$ (in electrostatic unit of charge, esu) and the mass of an electron $m_e$ (grams) are constants, and $n_e$ is the local electron number density per cubic centimeter.

\begin{align}
    \omega_{pe} = \sqrt{\frac{4\pi n_e e^2}{m_e}} = 8.98 \text{kHz} \sqrt{n_e/cm^3}
    \label{eqn_plasfreq}
\end{align}

Type II emissions can be classified by their emitted wavelength, ranging from metric (m), to decametric-hec\added{t}ometric (DH), all the way to kilometric (km).  Type II emissions can start at frequencies around 150 MHz, which corresponds to the plasma frequency a few solar radii from the Sun, and can drift past 1 AU where the local plasma frequency is \added{approximately }25 kHz.  \cite{Gopalswamy2005} found evidence that the kinetic energy of the CME seems to determine the lifetime of type II radio bursts, with the most energetic events having bursts that extend from the metric frequencies, down into the DH and kilometric ranges.  Less energetic CMEs on the other hand have no type II emission, or emission that stops in the metric frequencies, before the DH wavelength ranges.  Wide ranging surveys of events by \cite{Gopalswamy2008} found that basically all DH type II bursts are associated with SEP events at Earth if one accounts for magnetic connectivity.  The intensity of the associated SEP fluxes also increases with CME speed and width (i.e. kinetic energy).  Another recent broad ranging study by \cite{Winter_2015} further solidifies this relationship where it is reported that every single SEP event seen by Wind WAVES from 2010-2013 with a peak $>$ 10 MeV flux above 15 protons $cm^{-2} \cdot s^{-1} \cdot sr^{-1}$ is associated with a DH type II burst, most of which occur alongside a type III burst. 



With this general relationship cemented, the primary goal of this work is to better understand which plasma parameters are most important for type II burst generation within the context of CME structure and shock geometry.  To do this, a methodology is developed to take CME simulation data and create synthetic type II bursts for hypothesized burst locations across the entire simulation time.  Existing methodologies from those such as \cite{Reiner2007} for using type II radio data to profile the velocity and acceleration of a CME are extended in order to create a ``stencil", which is an estimated spectral time profile employing Gaussian intensity profile fitted over the observed burst intensity.  It is this Gaussian profile fit to the observed radio burst that the synthetic spectra are scored against.  These two methodologies are then used as cornerstones in a framework in which the following optimization problem is posed:``What progression of simulated burst locations provide a synthetic \replaced{spectra}{spectrum} that best matches a single spacecraft observed \replaced{spectra}{spectrum}?"  

Within this framework, one may track different features of the simulated CME to test different hypotheses of what plasma parameters are causal in creating the type II bursts around CMEs.  These burst locations in turn mark where the acceleration of energetic particles is happening as the CME expands throughout the heliosphere.  These burst locations can then be fed to energetic particle codes like EPREM \citep{Schwadron2010, Schwadron2014} that track the magnetic connectivity with Earth to forecast energetic particle flux at Earth, though that is not done in this work.  One promising long-term goal of applying this framework is finding a common set of thresholds of plasma parameters that create synthetic type II spectra that match observed spectra across many recreated simulations of major CMEs.  Such findings would be valuable information for constraining physical theories of particle acceleration at CMEs, and would also be immediately applicable to real time space weather operations for those at NOAA's Space Weather Prediction Center and others.  A set of key thresholds that mark the site of type II burst generation and particle acceleration would help enable more reliable space weather prediction, especially with the use of faster than real time simulations of incoming CMEs on the horizon.  

In this work, these methodologies and optimization framework are developed with the focus on a single case study in the form of the well documented 2005 May 13 CME.  A general outline of the remaining sections of the paper are as follows: in section 2, previous work in tracking CMEs and type II bursts is reviewed, with major steps in methodology and subsequent understanding being identified.  Section 3 describes the simulation of the CME by \cite{Manchester2014} (performed with the Alfv\'en Wave Solar Model (AWSoM) developed by \cite{van_der_Holst_2014}) that is used throughout the rest of the paper.  Section 4 develops the methodology to create an event specific ``stencil" of a type II burst from observed single spacecraft data. In this case, the methodology assumes CME velocity and solar wind density profiles to fit the evolution of the type II burst over time.  This approach yields, for each moment in time, an idealized Gaussian spectral profile around a central frequency, which is used to create the ``stencil" that synthetic spectra are scored on according to their similarity with the observed Wind/WAVES radio data.  Section 5 develops the methodology to create synthetic spectra from the simulation, using the upstream electron density to determine the observed frequency profile for each moment in time.  In section 6, the methodologies are applied to frame the optimization problem:``What progression of simulated burst locations provide a synthetic \replaced{spectra}{spectrum} that best matches the single spacecraft observed \replaced{spectra}{spectrum}?"  Here a range of CME model features based on different physical thresholds and geometrical factors are tested to identify the likely site of burst generation around the CME.  In section 7, possible sources of error and objections to the new methodologies are discussed and addressed to establish that the methodologies are appropriate for this optimization framework.  In section 8 the results from this investigation are discussed and future work is laid out.  

\section{Previous Work in Tracking CMEs \& Type II Bursts}


\subsection{Theories of Particle Acceleration}

Even with the general association of SEP events with CMEs and type IIs being well documented, the exact nature of the relationship is unknown in many respects.  One of the largest outstanding issues is identifying which part of the CME is accelerating the particles, and how it does so.  A variety of possible acceleration mechanisms have been proposed: one is diffusive shock acceleration at the CME-driven shock \citep{Tylka_2005, Tylka_2006a, Tylka_2006b}, second is SEP generation with CME expansion and acceleration in the low corona \citep{Schwadron_2015}, third is SEP acceleration occurring at a current sheet sun-ward of the ejected flux rope \citep{Forbes91, Forbes95}, a fourth possibility is non-local acceleration is occurring diffusely over a broad region as coronal plasma is diverted and compressed by the expanding filament \citep{Zank_2015} and a fifth possibility is acceleration at compression regions in the CME sheath \citep{Manchester_2005, Kozarev2013}.  If the location and conditions of SEP acceleration within the CME were better understood, it would advance the capability to forecast and predict the occurrence and intensity of energetic particles at Earth.  This is a very important problem, and much effort over the decades has gone into pinning down the origin of type II bursts within ICMEs.

\deleted{
Type II bursts were used early on to track the propagation of ICMEs, usually assuming that the shock location could be fit to the corresponding coronal location by matching its plasma frequency in studies like \cite{Pinter1982}.  Later on, \cite{Pinter1990} used type II emission data to develop a semi-empirical model of ICME trajectories and transit times to the Earth.  \cite{Reiner1998} showed that it was possible to roughly track the progression of a CME with the accompanying type II emission, using the spinning Wind/WAVES antenna to do basic direction finding in tracking the origin of the burst.  A series of papers by \cite{Corona2011, Corona2013, Corona2015} describe an analytical blast wave propagation model developed to describe the propagation process of a CME-driven shock.  The process starts with a period of roughly constant speed where the shock is actively driven.  It then goes under an intermediate stage of decoupling, after which the shock is undriven, and decelerates approaching the speed of the surrounding wind.  This theory draws on decades of work in blast wave modeling such as \cite{Cavaliere1976}.  The final results in \cite{Corona2015} show generally a good match for the shock model and the track in frequency space with some selected type II burst data.  
}

\deleted{
These modeling efforts were mirrored for type III bursts by studies such as \cite{Dulk1987} and \cite{Reiner2015}, which combined models of electron density with radio spectra to back out the speed and deceleration of the physical electron packets producing various type III bursts to great success.  \cite{Leblanc1998} used a curated collection of type III burst data with clear links to energetic electron and Langmuir wave measurements from spacecraft to derive an empirical model of the electron density profile from the frequency profiles of the bursts.    
}

\subsection{Advances in Space Based Observations}

A major methodological leap was achieved by \cite{leblanc2001} in combining LASCO white light coronagraph data with radio observations to trace the progression of CMEs and their associated shock waves in relation to their observed type II bursts.  They found that sometimes the inferred kinematics of the shock matched the observed frequency profile, but often there was some discrepancy, indicating a more complicated relationship.  \cite{Reiner2007} also applied kinematic models to reproduce the type II frequency drift from the propagation of ICMEs/shocks, using SOHO LASCO white-light observations to constrain the initial speed of the ICME.  \cite{Gopalswamy2009} later showed how STEREO coronagraph data from 1.4 - 4 $R_{\odot}$ can be used to observe CMEs at the onset of type II bursts.  

Another methodological leap was enabled by the STEREO spacecraft in the form of multi-point observations for both radio and white light coronagraph observations.  \cite{Reiner2009} used Wind/WAVES and STEREO/WAVES to give the first three-point measurements of type III bursts, constraining their beaming characteristics.   \cite{Martinez_Oliveros2012} utilized the same set of spacecraft to study a type II bust from an August 1, 2010 CME-CME interaction event.  They reported a spread in type II radio source positions, indicating an complex and extended structure.  \cite{Krupar2014} used both STEREO spacecraft to estimate multiple wave vector directions of type III bursts to provide estimates of their emitting regions and their corresponding source sizes. 

\cite{Makela2016} also utilized multi-point goniopolarimetric observations from STEREO/Waves and WIND/Waves receivers in addition to LASCO white light coronagraph data to triangulate the source of type II emission in an event of CME-CME interaction in DH wavelengths.  Through comparison of radio data and white light coronagraph data they found that scattering affects radio wave propagation at lower frequencies.  \replaced{The WAVES antennas indicated that source directions for frequencies near 1 MHz matched the location of the leading edge of the primary CME seen in the images of the LASCO/C3 coronagraph.}{Analysis of the radio data around 1 MHz from the Waves antennas for this event indicated that the source direction matched the front of the CME, which was imaged by the LASCO/C3 coronagraph.}   Studies like these that combine radio data with coronagraph images are powerful, but lack any way of observing many different plasma parameters like the magnetic field that may affect burst generation but cannot be remotely sensed.

\cite{Magdaleni14}, like \cite{Makela2016}, used multi-point observations to pin down the likely location of the burst utilizing recent techniques of goniopolarimetric observations from STEREO/Waves and WIND/Waves spacecraft data.  However, \cite{Magdaleni14} also used an Enlil simulation and coronagraph images to compare the height of the CME with the radio data and track the propagation of the shock wave further than coronagraph data alone can.  They found that combined observations of the March 5, 2012 event suggested that the southern flank of the CME was the source of the radio emission.  Remote observations combined with simulated recreations of the event in question can provide additional insight into the physical parameters at the possible site of burst generation that cannot be remotely sensed, such as the magnetic field strength.  However, \cite{Magdaleni14} does not go this far, and uses the model merely to double check the shock arrival time and the kinematics of the CME.  The following sections of this work will go further and use the simulation data to probe the plasma parameters at the hypothesized site of radio emission that would normally only be available as \textit{in situ} observations.  

Progress in tracking type II bursts into lower frequencies continues with the Sun Radio Interferometer Space Experiment (SunRISE) \citep{alibay2018, Hegedus2019-1}.  SunRISE is a recently funded NASA Heliophysics Explorer Mission of Opportunity that will send 6 smallsats each with a 5 m dipole to observe radio frequencies between 0.1 - 25 MHz, which promises to fill the gap in imaging data for type II and type III bursts in the lowest frequencies below the ionospheric cutoff.  As opposed to multi-spacecraft radio observations done in \cite{Magdaleni14} and \cite{Makela2016}, which use various direction finding techniques, the SunRISE receivers will be synchronized to enable true signal correlation and interferometry that has until this mission has been limited to ground based arrays.  

\subsection{Advances in Ground Based Radio Observations}

There have also been efforts to combine ground based radio observations with space based observations of CME structure.  Using radio observations from 164 - 435 MHz by the Nan{\c{c}}ay radio-heliograph \citep{Kerdraon1997}, a number of metric type II bursts have been compared to EUV and X-ray maps in studies such as \cite{Klein1999} and \cite{Zucca2014}.  These metric type II events have also been studied in combination with the innermost edges of coronagraphs in works like \cite{Maia2000}.  These studies, while scientifically valuable, do not observe the DH type II bursts further from the Sun that are associated with the more intense SEP events \citep{Gopalswamy2005}.  Recent progress has been made on this front with low frequency interferometers like LOFAR \citep{vanHaarlem:2013gi} and OVRO-LWA \citep{Eastwood2018, Chhabra2020} coming online.  \cite{Zucca2018} combined white light coronagraph data from SOHO and STEREO A along with radio interferometer data from LOFAR capturing type II burst harmonic emission between 50 - 70 MHz.  This was the first time concurrent radio positional observations could be made for the more energetic DH type II bursts.  They found that the emission originated from the northern flank of the expanding CME.  They employed a model named magneto-hydrodynamic around a sphere polytropic (MASP) developed in \cite{Linker1999}.  This model uses photospheric magnetograms from SDO/HMI to seed the simulation.  One drawback to their approach is that this model utilizes a rather simple energy equation that cannot accurately describe fast and slow solar wind streams.  It has also been reported by \cite{Rouillard2016} that this model can provide densities up to an order of magnitude too high.  This is the same vein of methodology employed in this paper, except this work employs a more sophisticated MHD model and forms an inverse problem to find burst locations using synthetic spectra instead of using direction finding or imaging.  The AWSoM model \citep{van_der_Holst_2014} used in this work, described more fully in the next section, uses a more realistic energy equation and more closely recreates observed plasma density \citep{Manchester2014}.    

In another study combining ground based and spaced based measurements, \cite{Morosan2019} used LOFAR radio imaging in combination with EUV and coronagraph imaging to piece together the relative location of radio emission on an energetic 2017 CME. They recorded radio emission coming from the southern flank of the CME, then minutes later at the northern flank, where shocks were thought to form in both flanks in regions of relatively low Alfv\'{e}n speed and high $\Theta_{Bn}$, where \added{$\Theta_{Bn}$ is the local angle between the shock normal and the magnetic field.  These} parameters were discerned by a relatively simple potential field source surface (PFSS) model that takes as input the observed photospheric field and outputs a model of the coronal magnetic field up to 2.5 solar radii. 

\subsection{Advances in MHD Simulations}

Another work that employs simulations to better understand CME observations is \cite{Kozarev2013}, which connects a BATSRUS MHD simulation of the same 2005 May 13 CME (also the focus in this work) to a SEP propagation code named EPREM from \cite{Schwadron2010}.  EPREM tracks the energization of particles, taking into account compressive acceleration around shocks and compression regions, as well as adiabatic cooling over the course of CME expansion.  They employ this model to study the SEP acceleration around the CME between 1.8 - 8 $R_{\odot}$, finding evidence of accelerated particle energies of 100 MeV, thought to originate in the sheath region right behind the shock.  This is largely in agreement with the findings of this study, which finds that a large swath of the simulated entropy derived shock overlaps to some degree with observed radio data, though this work goes further in attempting to pinpoint where in the shock this process is happening.

\added{ Another approach to understanding radio emission around CMEs is taken in a series of papers by \cite{Schmidt2012, Schmidt2012_2}, where an analytic bolt-on model is described that runs on top of any MHD simulation, calculates the energy transfer from Langmuir waves generated by an electron beam to the field, and outputs predicted radio fluxes seen from Earth.   This first analysis runs on top of an azimuthally symmetric 2.5D MHD model of a simple CME with no Carrington Rotation specific steady state pre-event solution.  They later expanded on this work in  \cite{Schmidt2013}, where a full 3D MHD model of a February 15 2011 CME was used and passed into the analytic type II burst algorithm.  There was good agreement of their simulated spectra with the recorded Wind/WAVES data, indicating that the framework described did a decent job at recreating the physics underlying the radio bursts.  However, their results were still a factor of $\sim 10-30\%$ off in frequency at some times, showing that there was still progress to be made.  There is also the question of unique solutions, was this calculation just one of many that yielded similarly good recreations of the data?  Applying this method across more events would help increase confidence in its results.  This is just what they did in \cite{Schmidt2014}, where they included analyses of MHD recreations of both the February 15 2011 CME and the March 7 2012 CME.  Again they found significant overlap with their calculated spectra and the real data.  In another more recent work \cite{Schmidt2016}, the group demonstrates the use of their bolt on model with an MHD recreation of the 2013 Nov 29 - Dec 1 CME.  They find good agreement between their model and the recorded STEREO/WAVES data, extending the comparisons out to 1 AU, with plasma frequencies below 100 kHz.  This series of papers is in a very similar spirit to this work, with a key difference being that this work assumes no single burst generation mechanism, and instead creates many synthetic spectra to score and compare to the observed radio data.  }

\subsection{Using \textit{in situ} Plasma Data}

There has already been some interest in determining which plasma parameters are important for generation of Langmuir waves, which act as {\it post facto} indicators of particle acceleration.  \cite{Pulupa2010} analyzed Wind data of {\it in situ} measurements of Langmuir waves upstream of interplanetary shocks to see which plasma parameters were correlated with Langmuir activity.  They found that the variable most correlated with Langmuir activity is the de Hoffmann-Teller speed.  The de Hoffmann-Teller frame, mathematically defined in a following section, is the reference frame where the convective electric field vanishes on both sides of the shock.  \replaced{The angle between the shock normal and the magnetic field, $\Theta_{Bn}$,}{$\Theta_{Bn}$} is also thought to be important in deciding the acceleration zone, and was also found to be highly correlated with Langmuir activity.  However, Langmuir activity was found to be correlated with many other variables as well, including the strength of the upstream and downstream magnetic fields, but no single parameter being decisively predictive.  Nevertheless, \cite{Pulupa2010} provides a useful starting point for what plasma parameters are likely to be successful in recreating \added{observed} radio spectra.

Studies like this one are useful for understanding which shock parameters may be important in SEP acceleration at CMEs, but they do not provide predictive capability since these plasma parameters are only observable \textit{in situ}, and only provide information about Langmuir waves as a marker for the SEP acceleration site at 1 AU.  A similar study can be done \textit{in silico}, where a sophisticated simulation can provide synthetic \textit{in situ} data over the entire simulated domain.  Notably, this means that one can observe the plasma parameters when the CME is within 20 solar radii of the Sun, when most of the SEP acceleration happens.  By combining this synthetic \textit{in situ} data with the actual observed radio data, one can frame the inverse problem of ``Which hypothesized progression of burst sites matches the observed frequency profile the best?"  Since the emitting frequency is only a function of upstream plasma density, \replaced{one can easily}{extraction of this information from an MHD simulation allows one } create a synthetic \replaced{spectra}{spectrum} for a given hypothesized source progression and compare it to the observed \replaced{spectra}{spectrum}.  Given a quantitative way to assign a similarity score to a synthetic \replaced{spectra}{spectrum}, one may directly compare hypothesized burst locations perhaps find a progression that nicely matches the observed frequency profile.  This scoring methodology is fleshed out in sections 4 and 5.  This way, one may directly test different combinations of physical thresholds and geometrical simulation features to determine what factors are causal in type II burst generation, and correspondingly, SEP acceleration.  Solving this inverse problem also bypasses the need for simulating the exact physical processes energizing particles creating the bursts, and directly tells us what areas of the CME are most likely to be the site of particle acceleration.  These locations may then be inspected more closely with the additional data from the simulation to see which plasma parameters are important for type II burst generation to occur.  In the following sections, these methodologies are applied to the 2005 May 13 radio-loud ICME to identify likely regions where the type II burst was generated, utilizing an AWSoM MHD recreation \citep{van_der_Holst_2014, Manchester2014}, and Wind/WAVES \citep{Bougeret1995} radio \replaced{spectra}{spectrum}. 


\section{Simulating the 2005 May 13 CME} 

The CME simulation (described in \cite{Manchester2014}) was performed using the Alfv\'{e}n Wave Solar Model (AWSoM) \citep{van_der_Holst_2014}, which models the time-average ambient solar wind over a Carrington Rotation (CR) \replaced{with the use of}{based on photospheric radial magnetic field measurements from} synoptic magnetograms. This level of detail is necessary for accurate simulations of upstream solar wind density as \cite{Manchester_2005} showed that the ambient solar wind structure strongly affects the evolution of CME structure.  \replaced{The CME is driven by a \cite{Gibson_1998} magnetic flux rope that possesses common characteristics of pre-event structures, including a dense helmet streamer with a cavity and dense core threaded by the flux rope.}{The simulated CME is driven by a \cite{Gibson_1998} magnetic flux rope that has some common pre-event characteristics, such as a dense core surrounding the flux rope and a dense helmet streamer with a cavity.}  This particular CME has been well studied, with works like \cite{Bisi2010} combining data from several suites of instruments to piece together a more complete picture of the event.  A SOHO C2 coronagraph showing the ``halo" CME for this event alongside its corresponding Wind/WAVES radio \replaced{spectra}{spectrum} is shown in Figure \ref{fig_coronagraph}.


\begin{figure*}[t]
\centering
\gridline{\fig{cme05.png}{0.4\textwidth}{SOHO C2 White Light Coronagraph}
\fig{allWindMedLine.png}{0.55\textwidth}{Wind/WAVES Radio \replaced{spectra}{spectrum}}}
\caption{\textmd{\textit{Left:} SOHO C2 coronagraph image of the 2005 May 13 ``halo" CME that is Earth-directed. \textit{Right:}  Wind/WAVES radio \replaced{spectra}{spectrum} accompanying the CME.  The highly intense emission in white is the type III burst, while the secondary red stripe above it is the type II burst, which is sustained for over \replaced{24 hours for this event.}{33 hours for this event, all the way out to 1 AU.}  A white dotted line has been plotted over the first $\sim 2$ hours of the type II burst for visibility.}}
\label{fig_coronagraph}
\end{figure*}

The AWSoM model, recently developed at the University of Michigan \citep{van_der_Holst_2014}, includes a phenomenological model of low-frequency Alfv\'{e}n wave turbulence to heat the corona and accelerate the solar wind.  At the inner boundary, the Alfv\'{e}n wave energy is injected with the Poynting flux proportional to the magnetic field strength.  \replaced{Counter-propagating waves are present on open field lines as a result of the outward propagating waves being  partially reflected by the Alfv\'{e}n speed gradients and field-aligned vorticity.}{Counter-propagating waves are present on open field lines due to Alfv\'{e}n speed gradients and field-aligned vorticity reflecting some of the outward propagating waves backwards.}  Wave energy enters both ends of closed field lines to produce near-balanced turbulence at the apex of the loops. AWSoM currently has 3 temperatures, an isotropic electron temperature, and separate parallel and perpendicular ion temperatures, though the results here were obtained from a 2 temperature simulation, one for electrons and a second for isotropic ions.  To distribute the heating of turbulence dissipation across the different temperatures, the model uses the linear wave theory and nonlinear stochastic heating as presented in \cite{Chandran_2011}. AWSoM also incorporates both collisional and collisionless heat conduction into its treatment of electron thermodynamics.    This all leads to a more realistic simulation that better matches real data \citep{Sachdeva_2019}.  

\begin{figure*}[t]
\centering
\includegraphics[width=0.9\textwidth]{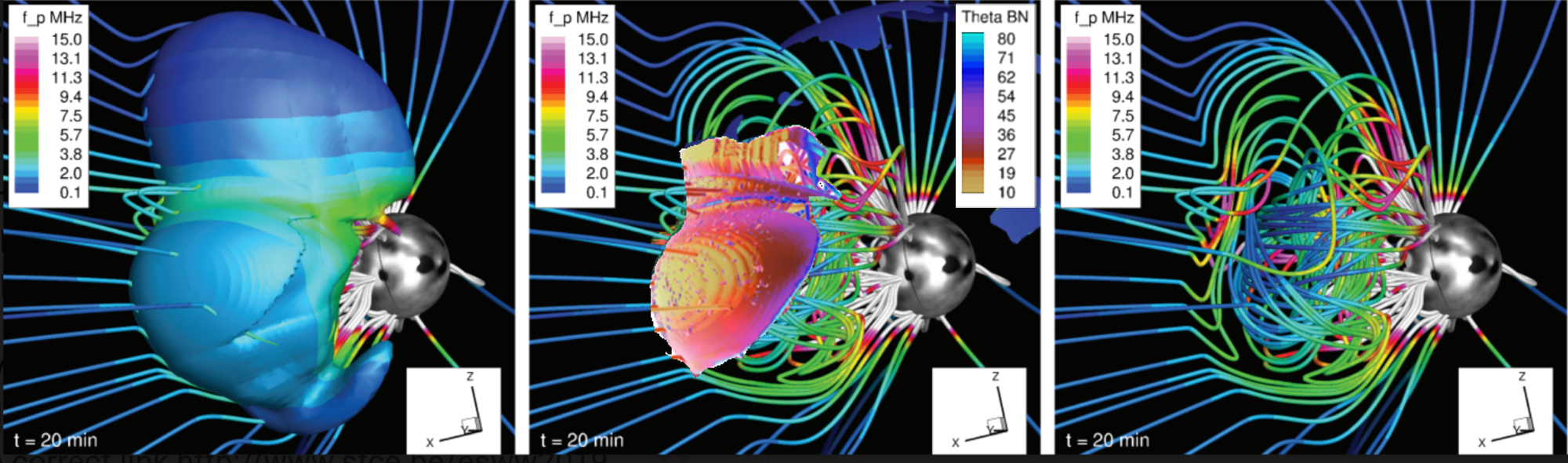}
\caption{\textmd{One frame from the AWSoM MHD Simulation the radio-loud 2005 May 13 CME 20 minutes after initiation. The left-hand panel shows the shock surface (defined by $\geq$3.5x density compression ratio) colored to show the plasma frequency, while the central panel shows the shock surface defined by a $\geq$4x entropy enhancement colored to show the shock angle $\Theta_{Bn}$.  The right hand panel shows only the magnetic field configuration.  In all cases, field lines are colored to show plasma frequency.}   
}
\label{fig_mhdsims2}
\end{figure*}

Results from the simulation of the 2005 May 13 CME event are seen in Figure \ref{fig_mhdsims2}. Here, different physical quantities relevant to SEP/acceleration and radio emission are shown 20 minutes after initiation. The panel on the right shows the magnetic flux rope being expelled from the corona with a peak speed of nearly 2000 km/s, which closely matches the speed fit by \cite{Reiner2007} for this same event.  The left panel shows the full extent of the region of the simulation with \replaced{at least 3.5x density compared to before the event to identify the density derived shock}{local densities that are at least 3.5x the pre-eruption values to identify the density derived shock}, colored to show the upstream plasma frequency.  The central panel shows a region of strongly-shocked plasma\replaced{with an iso-surface of 4x entropy enhancement colored to show the shock angle $\Theta_{Bn}$.}{ identified by selecting regions of the simulation with entropy values that are at least 4x the pre-eruption values, with the surface colored to show the shock angle $\Theta_{Bn}$.  Entropy is defined as $S = \frac{p}{n^{5/3}}$ for plasma number density n and pressure p.} The right hand panel shows only the magnetic field configuration.  In all panels, the magnetic field lines are colored to show the plasma frequency.   

Data from MHD simulations contains a wealth of information that would be impossible to get at similar cadence and resolution with \textit{in situ} or remote sensing data.  Each data point contains cell averaged quantities of magnetic field, plasma density, momentum, and internal energy.  These parameters may then be used to calculate second order parameters like entropy, de Hoffmann-Teller Frame velocity, angle between shock normal and upstream magnetic field $\Theta_{Bn}$, and more.  Some of these derived plasma quantities have been found to be highly correlated with Langmuir waves upstream of shocks from \textit{in situ} observations \citep{Pulupa2010}.  Model data extractions (iso-surfaces and field lines)  using these plasma parameters along with geometrical sub-selections will be used in the following sections to test different hypotheses of burst locations by comparing their synthetic spectra with the observed Wind/WAVES \replaced{spectra}{spectrum}.  

\section{Methodology for Creating a Scoring Stencil}

Synthetic spectra do not have physical units such as $V^2$ or spectral brightness like spacecraft data, as they are not made by directly simulating the physics of burst generation.  Briefly put, a synthetic \replaced{spectra}{spectrum} is created by taking a suspected source region from \replaced{a}{an} MHD simulation, and calculating the upstream plasma frequency for each point, and plotting the resulting distribution in a histogram for each time step, with the frequency bins being the same as the Wind/WAVES data.  So it is not feasible to directly compare synthetic spectra with spacecraft observed spectra, as they have dimensionless units, consisting only of raw counts of simulation points with a certain plasma frequency, compared to power received at an antenna in a given frequency.  Thus it is necessary to create a custom ``stencil", i.e, a distribution over frequency and time derived from the observed data, so one may analytically assign a similarity score between the synthetic \replaced{spectra}{spectrum} and the observed \replaced{spectra}{spectrum}, using the stencil as a proxy.  This methodology gives one the flexibility to be agnostic to the actual physics generating the type II burst, assuming only that the emission frequency is the upstream plasma frequency.  This forms the basis of a framework where one can find test hypothetical burst locations by quantifying how well \replaced{their}{the formed} synthetic spectra match the stencil.  

Note that this framework would not be possible without the assumption of the emission frequency being upstream plasma frequency.  It has been suggested by \cite{Bastian2007} that some events have type II like emission that comes not from the plasma emission mechanism, but from incoherent synchrotron radiation from near-relativistic electrons around the CME or its sheath.  If this were the case, one can potentially back out the distribution of energetic elections, but probably not constrain their location, since the emitting frequency is not tied to a location like it is for the plasma emission mechanism, where the upstream plasma frequency is affected by the distance from the Sun.

\begin{figure*}[h]
\centering
\gridline{\fig{longwindSpeedProfilesHarmonic.png}{0.9\textwidth}{}}
\gridline{\fig{windSpeedProfilesHarmonicShort.png}{0.9\textwidth}{}}
\caption{\textmd{Wind/WAVES radio \replaced{spectra}{spectrum} of the 2005 May 13 CME overplotted with the predicted center frequency of the type II burst from the CME kinematics from \cite{Reiner2007} following an $r^{-2}$ electron density model with a 1 AU density of 11 $n_e /cm^3$.  The first 2 hours from 17:00 onward are mainly emission at the fundamental plasma frequency, while the rest of the time the harmonic emission is dominant, with little overlap.  Top: Long term view of the the event.  Around the 50 hour mark on this plot, at 02:13 UT 2005 May 15 one can see \replaced{when the shock arrived at L1.}{the enhancement in radio emission that coincides with \textit{in situ} observations of the shock arrival by Wind and ACE \cite{Bisi2010}.}  Bottom: Short term view of the beginning of the event.  }}
\label{fig_harmonics}
\end{figure*}

\begin{figure*}[h]
\centering
\gridline{\fig{windSpeedProfiles.png}{0.9\textwidth}{}}
\gridline{\fig{stencilLongNorm.png}{0.9\textwidth}{}}
\caption{\textmd{Wind/WAVES radio \replaced{spectra}{spectrum} of the 2005 May 13 CME.  Top: Type II burst identified with three predicted frequency profiles, marking the upper, center, and lower frequencies of the stencil.  Each profile has the same kinematics as the curve in Figure \ref{fig_harmonics}.  Bottom: Stencil of the type II radio burst associated with the 2005 May 13 CME derived from above.  The dashed lines shared indicate the three sigma cutoff of the stencil, marking the bounds of the type II burst.}}
\label{fig_windData}
\end{figure*}

\added{
\begin{figure*}[h]
\centering
\includegraphics[width=0.9\textwidth]{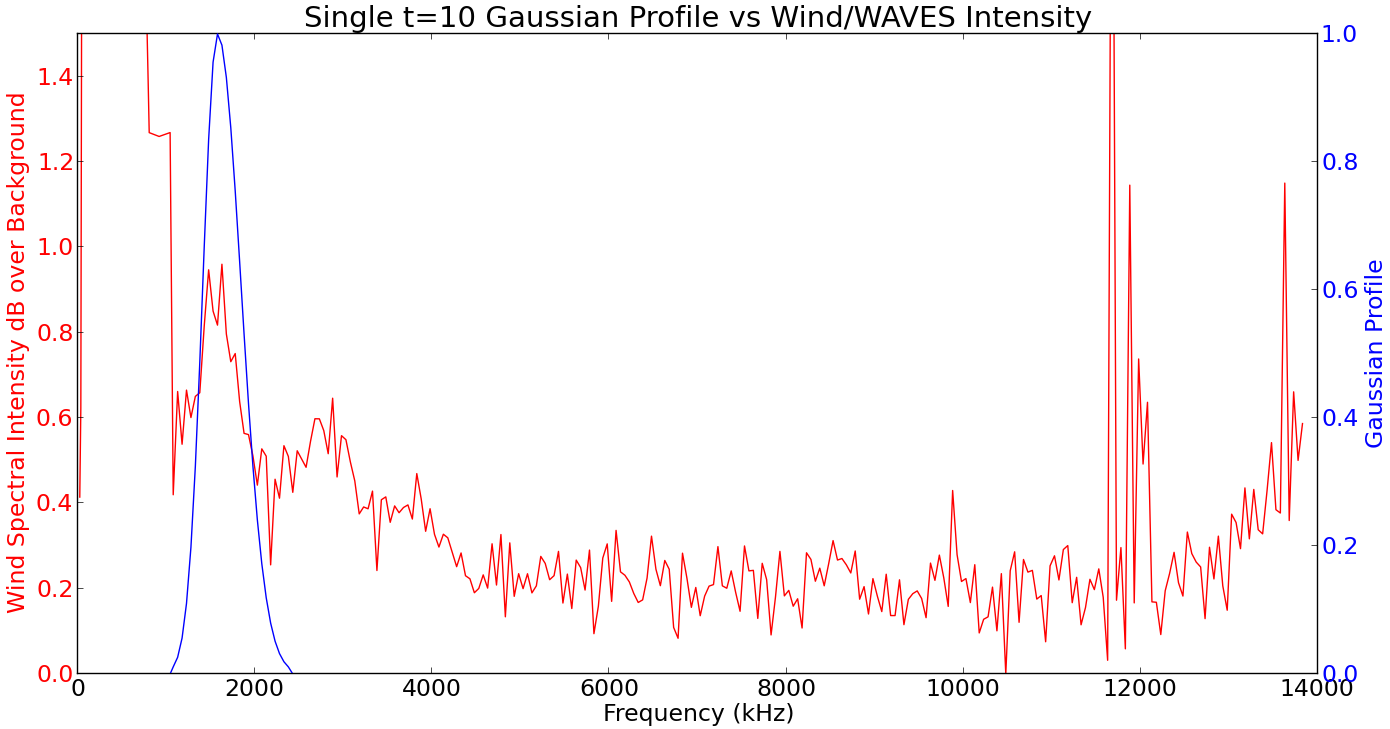}
\caption{\textmd{Wind/WAVES radio spectrum for 17:10, 2005 May 13 over-plotted with the Gaussian scoring stencil for that same time.  The type II burst is far from the brightest feature in the data, requiring manual partitioning to isolate.}}
\label{fig_singleSpectrum}
\end{figure*}
}

Since every CME is unique, having a different geometry and speed profile, each event requires a custom stencil to be created that synthetic spectra can be compared to.  In the current literature, there are examples of using basic parameterizations of type III bursts to fit the underlying kinematics of the electron packet driving the burst.  \cite{Dulk1987} simply fit a constant speed of the outgoing energetic electrons, while \cite{Reiner2015} included a constant deceleration term alongside an electron density and Parker spiral model to accurately capture the radio wave travel times.  They argue that the additional degree of freedom allows their model to better fit the frequency drift from spacecraft observed \replaced{spectra}{spectrum} as well as match the arrival time of Langmuir waves or local emissions.  \cite{Reiner2007} characterized the kinematics of CMEs from type II frequency drift, using SOHO LASCO white-light observations to constrain the initial speed of the ICME.  One of the events they analyzed was actually the same 2005 May 13 event studied in this work.  Here, this methodology is extended to create a physically motivated stencil based on CME kinematics.

The initial kinematics of the 2005 May 13 CME were derived from their recorded times in which the halo CME was detected with SOHO LASCO C2 (17:22 UT) and C3 (17:42).  These times correspond to an initial speed of 1690 km/s and an extrapolated liftoff at 16:47.  This initial speed however is not the true speed; it is only a plane of sky projected speed.  Since the event in question is a halo CME, the true speed is higher than the plane of sky estimate.  The event was detected  \textit{in situ} by the \textit{Wind} spacecraft at 1 AU on 02:13:30 UT on 2005 May 15, implying an average speed of 1246 km/s.  In addition, \cite{Reiner2007} found the measured radial shock speed at 1 AU to be 778 km/s.  This large deceleration between liftoff and 1 AU is expected for fast CMEs; \cite{Woo1985, Sheeley_1999, Manchester_2004} have shown that this deceleration increases with the initial speed of the CME, explaining the rather narrow range in speeds of shocks observed at 1 AU.

As \cite{Reiner2007} describes, there are many possible simple kinematic solutions satisfying these constraints, with higher initial speeds being matched with quicker deceleration for shorter periods of time before settling at 778 km/s at 1 AU.  They also note that the structure of the type II emission may be used to find the ``true" velocity profile.  \replaced{The value they came up with that best fits the center of the emitting frequency or harmonic was a $v_0 = 2500$ km/s, a deceleration of 26.7 m/s/s, and a deceleration time of 17.9 hours.  To go from this velocity model to a radio spectra, one needs to assume a solar wind density model.  \cite{Reiner2007} opts for a simple $r^{-2}$ density model determined by a prescribed density at 1 AU.}{By assuming a solar wind density model, one can convert a kinematic velocity profile for a CME-driven shock to a  predicted radio profile over frequency and time.  \cite{Reiner2007} opted for a simple $r^{-2}$ density model determined by a prescribed density at 1 AU.  They then adjusted the parameters of the kinematic profile until they found a good fit by eye to the center of the apparent type II event as the fundamental or harmonic emission.  This kinematic model had $v_0 = 2500$ km/s, a deceleration of 26.7 m/s/s, and a deceleration time of 17.9 hours.  The frequency profile described by this solution will be central in our own scoring stencil below.}    With this kinematic solution, one can see in Figure \ref{fig_harmonics} that in the first 2 hours the type II emission is primarily at the fundamental plasma frequency, while after 19:30 or so the harmonic emission is dominant, continuing for another \replaced{24 hours or so}{31 hours before the shock arrives at 1 AU, coinciding with \textit{in situ} observations of the shock by multiple spacecraft like Wind and ACE \cite{Bisi2010}.  This arrival of the shock may also be seen in the Wind/WAVES data as an enhancement in the type II burst at around 50 hours on the bottom panel of Figure \ref{fig_harmonics}}. 




One may also extend this characterization of the emission, fitting by eye a customized envelope created by manually choosing three time-frequency profiles using this quadratic model to partition the initial fundamental emission from 17:00-19:00.  This can be seen in Figure \ref{fig_windData} a, where the 3 lines surround the strongest part of the primary type II emission.  The center frequency line \replaced{(in logarithmic frequency scale)}{from the \cite{Reiner2007} fit} is used as the target frequency for each time step, with the lines surrounding it used to impose a Gaussian profile over the observed burst.  This Gaussian profile is normalized so that the middle frequency has weight 1.  For every moment in time, each half of the identified emission in frequency space is given a different sigma value to determine its frequency width, and is defined with 3 sigma being the upper/lower end of the by-eye fit of the data.  The simulation data used for this work is constrained to the first 2 hours of the event, which coincides with the period of the event where the fundamental emission is dominating.  Further studies would require a straightforward extension to the stencil to include the harmonic frequencies.  

The resulting Gaussian profile over frequency and time is shown in Figure \ref{fig_windData} b, and can be thought of as a stencil to which a synthetic \replaced{spectra}{spectrum} from an MHD simulation may be compared to and scored from, enabling a template matching technique that may be used to solve the inverse problem of finding the site of type II burst generation.  The stencil is dimensionless, having no physical units, and thus may be directly compared to synthetic spectra if they have the same time and frequency structure.  \added{An example of the Wind/WAVES spectrum over-plotted with the described Gaussian stencil is seen in Figure \ref{fig_singleSpectrum} for 10 minutes into the type II event at 17:10.  One can see in this figure that the type II burst is far from the brightest feature in the noisy data, making manual partitioning necessary for the creation of the stencil.  }A global element-wise \replaced{multiply}{multiplication} between the synthetic \replaced{spectra}{spectrum} and stencil followed by a global sum of all the channels over frequency and time will yield a similarity score.  This similarity score is linearly proportional to the fractional amount of data points in a simulated CME feature whose upstream plasma frequency in its synthetic \replaced{spectra}{spectrum} overlaps with the scoring stencil.  The generation of these synthetic spectra will be further explained in the following section.



\section{Methodology for Creating Synthetic Spectra}
The primary goal of this work is to better understand which plasma parameters are most important for type II burst generation within the context of CME structure and shock geometry.  Making this connection will lead to greater understanding of how gradual SEP events are created by CMEs.  By employing state of the art AWSoM simulations, one can recreate historical major space weather events.  Sophisticated simulations can provide synthetic \textit{in situ} data over the entire simulated domain.  Notably, this means that one can derive the plasma parameters when the CME is within 20 solar radii of the Sun, when most of the SEP acceleration happens.  By taking different data extractions of the simulation based on physical thresholds for different physical processes or based on CME geometry, synthetic spectra can be made and compared with the actual radio data. This radio data needs to come from space, since the lowest parts of the type II and III bursts are emitting below the Earth's ionospheric cutoff.  By matching synthetic spectra to the spacecraft-observed data, one may attempt to determine which physical thresholds must be exceeded in order for burst generation to occur.  Confirmation of similar physical thresholds applied to synthetic spectra that generally match observations across multiple events would pin down the specific physics triggering the burst generation as well as the location of particle acceleration.  This work is focused on a single radio loud event from 2005 May 13 at 16:47 as a test case for this new framework.    




To this end, the first 2 hours of an MHD simulated recreation of the radio loud CME is compared with the Wind/WAVES radio \replaced{spectra}{spectrum} to identify likely regions where the type II bursts are generated.  Data extractions from the simulation are made based on combinations of physical thresholds, as well as CME/shock geometry, including solar longitude and latitude.  From these data, synthetic radio spectra are made and compared to the stencil described in the previous section.  


\begin{figure*}[b]
\centering
\gridline{\fig{2dhistTimeFreqFrontBit0Sum.png}{0.45\textwidth}{}
\fig{interpSyntheticFrontBit0.png}{0.55\textwidth}{}}

\caption{\textmd{Synthetic spectra from the MHD simulated entropy derived shock.  Left: Uninterpolated 2D histogram for a sub-selection of the shock nose.  Right: Synthetic \replaced{spectra}{spectrum} interpolated to the 1 minute cadence of Wind/WAVES.  In both plots, the columns are sum normalized to 1.0 for a maximum scoring potential of 1.0/minute.}}
\label{fig_synthBit}
\end{figure*}

Subsets are extracted from the 3D data files of the AWSoM simulation, typically iso-surfaces that meet some physical criteria.  These data files contain some basic plasma parameters that can be used to calculate secondary plasma parameters at the included simulation points, including the upstream plasma density used to calculate the upstream plasma frequency.  The \replaced{three}{four} main data types used here: iso-surfaces where points with at least a 3.5 MK temperature used to identify the CME current sheet, shocked plasma with a \added{density }compression factor of 3.5 or more, and points where the entropy changed by a factor of 4 or more, indicating the strongest (highest Mach number) shock.  Each of these data features is produced with a time cadence of 2 minutes for the first 40 minutes of the simulation, and later to transitioning to every 5 minutes producing a total of 35 data files over 120 minutes.  Two hours of simulation time brings the CME out to 20 solar radii, corresponding to a local upstream plasma frequency around 300 kHz.  \added{The one exception to this simulation sampling cadence is the selection for Fast Magnetosonic Mach $\# \geq 3$ sub-selection, which due to a lack of resources only combines the six times of [10, 20, 30, 40, 60, 90] minutes after the start of the simulation.  This results is a coarser data set that may still be used to inquire into the usefulness of this parameter, defined as the ratio of the shock speed to the fast magnetosonic speed, which has been shown to be associated with energetic particle acceleration at values $> 3$ \cite{Rouillard2016}.  }With these broad data sets in hand, finer extractions can then be applied based on plasma parameters considered to be important for SEP acceleration such as upstream \replaced{total velocity}{velocity magnitude} and de Hoffmann-Teller frame velocity \citep{Pulupa2010}, as well as subsets based on the geometry and spatial location of the data.  Synthetic spectra are created by calculating the upstream plasma frequency for each point and then plotting the resulting distribution in a histogram.  \replaced{We do this}{This is done} for each time step using the same frequency bins as the Wind/WAVES data.  For every time step of data saved, some preliminary processing must be done.  Depending on the simulation domain, different coordinate transformations must be performed.

The Solar Corona (SC) component of the MHD simulation goes out to 20 $R_{\odot}$ and is done in a rotating reference frame; SWMF calls it Heliographic Rotation (HGR), while SunPy calls it Heliographic Carrington (HGC).  The Inner Heliosphere (IH) simulation domain starts at 20 $R_{\odot}$ and can go out to several AU.  The IH domain is done in an inertial reference frame; SWMF calls it Heliographic Inertial (HGI), while SunPy calls it Heliocentric Inertial (HCI).  Data from both of these domains are transformed into the Heliocentric Earth Equatorial (HEEQ) frame, where the X axis is the Sun-Earth line, and Z is the Solar north pole.  These coordinate systems are described fully in \cite{Thompson2006}.

The HEEQ data for each time step is then sorted by \textit{x} coordinate, and a routine is applied to cull any low frequency points that appear to be behind higher frequency points from the Earth’s perspective.  This is done efficiently by employing fast algorithms for global nearest neighbor calculations.  Starting with the largest \textit{x} coordinate (closest to Earth), the plasma frequency is checked, then compared to any Sunward neighbors within a given angle on the sky, chosen here to be .001 \replaced{Rs}{solar radii} \added{ from the vantage of Wind at L1}, corresponding to 1 arc second in the plane-of-sky from Earth.  Any neighboring data points within this obscuration angle with a lower plasma frequency value than the Earthward neighbor is culled.  This reflects the physical fact of absorption of these lower frequency waves when passing through higher density areas.  Wind/WAVES was parked at L1 in the Sun-Earth line during this time period, so this line of sight culling is best done in this HEEQ frame to match the observed spectral data.  Secondary scattering effects that can enable lower frequency emission to take non-line of sight paths around higher density areas that would normally absorb them are not considered in this work.  The remaining points can then be further reduced using customized physical cutoffs to investigate the resulting distribution in upstream plasma frequency.  \added{There is also a special correction done on the entropy sub-selection to remove artifacts in the data that are much closer to the Sun than the rest of the extracted data.  These points have upstream magnetic field magnitudes on the order of 1 Gauss, as opposed to the rest of the data with values of less than 0.1 Gauss, and constitute around $1\%$ of the total data in the sub-selection.  This correction is done for all figures using the entropy sub-selection in the paper.} 

\replaced{For each time step, a histogram is created}{To form the base of a synthetic spectrum, a histogram is created for each time step} with the same frequency bins as the Wind data.  This histogram makes a single column of the synthetic \replaced{spectra}{spectrum}.  The entire synthetic \replaced{spectra}{spectrum} is a 2-D histogram, with each column sum normalized to 1.  This synthetic \replaced{spectra}{spectrum} is then interpolated to the same time cadence as the underlying data (1 minute for Wind/WAVES), while still enforcing the column sum normalization.  An example of this interpolation process is shown in Figure \ref{fig_synthBit}.  Then by employing an element-wise multiplication between the stencil and the interpolated synthetic \replaced{spectra}{spectrum}, and then globally summing the resulting array, one receives a score for how well the synthetic \replaced{spectra}{spectrum} from the selected simulation feature matches the spacecraft-observed data.  Since the column sum of the synthetic \replaced{spectra}{spectrum} is 1.0, and the stencil maximum is normalized to 1.0, a maximum score of 1.0/time step is achievable.  The current system most rewards synthetic spectra that match the center frequency, with a Gaussian profile only marginally rewarding close by frequencies, but a constant 1.0 weight can also be applied throughout the entire stencil envelope for a less picky algorithm.  Synthetic spectra for the strong shock region are shown in Figure \ref{fig_synthBit}.  The left panel shows the synthetic \replaced{spectra}{spectrum} as an uninterpolated 2D histogram, while the right panel shows the synthetic \replaced{spectra}{spectrum} interpolated to Wind/WAVE cadences, plotted over the stencil envelope.  The similarity score is seen in the title of this right panel, here a 29.26 out of a possible of 107.  The maximum possible score is 107 because there is only 2 hours, or 120 minutes of simulation data, but the simulation starts at 16:47, and the type II stencil starts at 17:00, leaving a maximum overlap of 107 minutes.  The slight distortion in the synthetic \replaced{spectra}{spectrum} around 1 MHz is due to the \replaced{transition of the Wind data from the RAD1 receiver }{transition from the Wind RAD1 receiver } below 1 MHz \replaced{which}{that} has a frequency resolution of 4 kHz to the RAD2 receiver above 1 MHz which has a frequency resolution of 50 kHz.  When plotting these sets of frequency channels with different resolutions, an apparent distortion at their interface near 1 MHz can be seen where a greater portion of points per frequency channel are seen above 1 MHz with the increase in frequency bin width.

\section{Finding the Type II Burst Locations}

By analyzing the plasma parameters in detail from an MHD simulation of a radio-loud event, one may attempt to solve the inverse problem of determining the location of the burst generation site.  One may extract different subsets from the simulated CME to create time-varying synthetic radio spectra using the upstream plasma density, and compare these to actual Wind/WAVES radio data to determine which hypothesized source region best matches the observed spacecraft data.  This bypasses the need for simulating the exact physics that are creating the bursts, and tells us what areas of the CME are most likely to be the site of particle acceleration.  These locations may then be inspected more closely with the additional data from the simulation to see which plasma parameters are important for type II burst generation to occur.  


\begin{figure*}[b]
\centering

\gridline{\fig{allWindOutlineShort.png}{0.5\textwidth}{Wind/WAVES Observed Radio \replaced{spectra}{spectrum}}
}

\gridline{\fig{interpSyntheticMach0.png}{0.5\textwidth}{Magnetosonic Mach $\# \geq 3$ Region}
        \fig{interpSyntheticFlankBit0.png}{0.5\textwidth}{Density x 3.5 Enhanced Region}
}

\gridline{\fig{interpSyntheticFront100th.png}{0.5\textwidth}{Entropy x 4 Enhanced Region}
        \fig{interpSyntheticSheet0.png}{0.5\textwidth}{Temperature $>$ 3.5 MK (Current Sheet)}
}

\caption{\textmd{Synthetic spectra generated from iso-surfaces extracted near the shock front of the 2005 May 13 CME.  Top: \deleted{Observed }Wind/WAVES \replaced{spectra}{spectrum}, with the envelope of the stencil over-plotted to show how the fundamental emission is accurately captured by the stencil.  \added{Upper Left: Synthetic spectrum made from the simulation regions with a magnetosonic mach $\# \geq 3$.  }Upper right: Synthetic \replaced{spectra}{spectrum} made from the simulation with density enhanced ($\times$ 3.5) sub-selection.  Lower left: Synthetic \replaced{spectra}{spectrum} made from the Entropy enhanced ($\times$ 4.0) sub-selection.  Lower Right: Synthetic \replaced{spectra}{spectrum} made from the Temperature $>$ 3.5 MK sub-selection.  All \deleted{3 of these} synthetic spectra have in their title the calculated similarity score to the stencil, and also have the outer envelope of the stencil plotted in white dashed lines.  The entropy enhancement sub-selection has the highest overlap with the stencil, reflected by the fact that its similarity score is the highest of the \replaced{three}{four, even after taking into consideration the shorter time period of the Mach $\#$ sub-selection}.  The slight distortion in the synthetic spectra around 1 MHz is due to the \replaced{transition of the Wind data from the RAD1 receiver }{transition from the Wind RAD1 receiver }below 1 MHz that has a frequency resolution of 4 kHz to the RAD2 receiver above 1 MHz that has a larger frequency resolution of 50 kHz, which captures a greater portion of points per frequency bin.}}
\label{fig_synspectra}
\end{figure*}

It is known from \cite{Aguilar05} statistical analysis of type II bursts that these bursts have a small bandwidth to frequency to ratio (BFR) of between 0.1 - 0.4 in the frequency range 0.03 to 14 MHz of the RAD1 and RAD2 antennas. However, \replaced{simulations of CME shocks have shown that if the entire shock surface was emitting a Type II burst, the frequency spread would be far larger.}{the MHD recreation of this 2005 May 15 event shows if the entire density enhanced shock was emitting a Type II burst, the frequency spread would be far larger.}  This effect can be seen in Figure \ref{fig_mhdsims2} left and Figure \ref{fig_synspectra} top right where the density enhanced region from the CME has upstream plasma frequencies spanning over an order of magnitude.  In this section, thresholds of plasma parameters will be used to identify and track smaller parts of the CME through space and time to create synthetic spectra that can better match the spacecraft observed radio \replaced{spectra}{spectrum}.  

\subsection{CME data extractions}

The starting point for this analysis are the \replaced{three}{four} sub-selections of data points from the CME simulation: first, points where the temperature lies at 3.5 MK, which occurs at the CME current sheet, second, points that are shock compressed to a factor of 3.5 in their density compared to the ambient level, \deleted{and }third, points that have their entropy changed by a factor of 4 or more\added{, and fourth, points that have a Fast Magnetosonic Mach $\# \geq 3$}.  These data, extracted as iso-surfaces from the 3D domain, are coming from the first 2 hours of simulation, with a time cadence of 2 minutes in the beginning of the run, transitioning to every 5 minutes later in the run for a total of 35 time points over 120 minutes.  The extent of the simulation brings us out to 20 solar radii, corresponding to a local upstream plasma frequency around 300 kHz.  \added{The one exception to this simulation sampling cadence is the selection for Fast Magnetosonic Mach $\# \geq 3$ sub-selection, which due to a lack of resources only combines the six times of [10, 20, 30, 40, 60, 90] minutes after the start of the simulation.  This results is a coarser data set that may still be used to inquire into the usefulness of this parameter.  }With these data extractions in hand, finer selections can then be applied based on plasma parameters thought to be important for SEP acceleration, or sub-selections based on the geometry of the CME.  








Figure \ref{fig_synspectra} shows the synthetic spectra from these \replaced{three}{four} model source regions alongside the actual radio data in the \replaced{upper left}{top} panel along with the stencil envelope.  One can see that the \added{bottom left panel with the entropy-derived }shock front matches the actual frequency \replaced{spectra}{spectrum} best, \replaced{matching}{aligning with} the gradual downward curve in frequency over time.  The current sheet \replaced{spectra}{spectrum in the lower right} stays mostly in the high frequency portion of the figure, reflecting the fact that the current sheet stayed close in to the sun even as the flux rope driving the shock moved outwards.  The shape of this emission roughly matches that of a solar type IV radio noise storm \citep{Takakura1963}.  This may be proof that these radio storms originate from the current sheet formed behind a large solar shock wave.  The synthetic \replaced{spectra}{spectrum on the upper right} for the density enhanced region shows the correct general shape of the type II burst, but is far too wide when compared to the relatively thin bandwidth-frequency ratio (BFR) of the actual \replaced{spectra}{spectrum}.  This could be a problem with the shock compression region being too inclusive, but it does mostly overlap with the correct spectrum.  \added{The synthetic spectrum on the upper left for the sub-selection based off Fast Magnetosonic Mach $\# \geq 3$ also shows the correct general shape, and is more focused around the stencil than the density enhancement sub-selection.  However, the data is not as closely aligned with the stencil as the entropy derived data.  This sub-selection criterion will be a promising parameter to check on in future studies, but will not be the focus of the remainder of the paper.}   

In the following subsections further restrictions are applied to the entropy derived shock data, as it has the highest initial overlap with the type II stencil.  This choice between the \replaced{three}{four} initial sub-selections is easy to make just by eye, but one can also look at the similarity scores between the sub-selections to see quantitative evidence that the entropy shock sub-selection has the highest level of overlap with the observed type II burst data\added{, even after taking into consideration the shorter time period of the Mach $\#$ sub-selection}.  Already one can see from Figure \ref{fig_synspectra} that looking at entropy enhancements could be a part of a set of key thresholds that can be used to predict where type II burst generation and particle acceleration is happening in CME simulations, possibly leading to improvements in space weather forecasting.

When analyzing the speed of this entropy identified shock region, it is found that the outermost extent of the shock goes from 3.92 $R_{\odot}$ at 8 minutes, to 23.99 $R_{\odot}$ at 120 minutes.  This corresponds to an average speed of about 2080 km/s, which is close to the expected \added{average }speed from the \cite{Reiner2007} fit of the CME kinematics from the frequency drift, which predicts an average speed of about 2400 km/s for this time period.  \replaced{The left panel of Figure \ref{fig_shockSpeed} shows the maximum and mean radial distances in the entropy enhanced feature for each moment in simulation time, and their respective inferred shock speeds.  The right panel shows mean velocity of the simulated shock, with the $V_x$ curve mostly matching the mean speed curve in the left panel.  This is expected, as this event was primarily Earth directed.  This realistic velocity profile of the simulated entropy derived shock front lends further confidence to the accuracy of the simulation results.}{The fit velocity profile of the simulated entropy derived shock is also decelerating, with an initial speed of over 2500 km/s, lending further confidence to the simulation's results.}

\deleted{By looking at the magnetic field strength of the entropy enhanced region, one sees that there are artifacts behind the bulk population of the shock that are artifacts from the data collection algorithm, capturing parts of the simulation domain where the shock has already affected the surrounding plasma.  These artifacts have field strengths over an order of magnitude larger than everything else from being so much closer to the Sun than everything else in the selected entropy enhanced feature, so a data sub-selection is employed to remove them at the beginning of every time step for each synthetic spectrum and variable contour graph below.  A variable contour graph of the magnetic field magnitude before this sub-selection is shown in Figure 9.  The difference this makes between synthetic spectra is negligible, as can be seen between the totally uncut entropy enhanced data in Figure \ref{fig_synspectra}, and the post magnetic magnitude sub-selection in Figure \ref{fig_frontSpectras}.  Visually the spectra look quite similar, and their similarity scores are only off by a score of 0.62, a relative difference of about $2\%$.  This has marginal impact on the generated synthetic spectra, but will improve the quality of maps of contour graphs generated with plasma parameters across the shock surface. }

\begin{figure*}[t]
\centering

\gridline{\fig{interpSyntheticFront100th.png}{0.45\textwidth}{All (non-artifact) entropy enhanced data included}
        \fig{interpSyntheticFront70th.png}{0.45\textwidth}{Only data points with distance $>$ 0.7*\textit{maxr} included }
}

\gridline{\fig{interpSyntheticFront80th.png}{0.45\textwidth}{Only data points with distance $>$ 0.8*\textit{maxr} included}
        \fig{interpSyntheticFront90th.png}{0.45\textwidth}{Only data points with distance $>$ 0.9*\textit{maxr} included}
}

\caption{\textmd{Synthetic spectra generated from subsections of the \added{entropy derived }shock front of the 2005 May 13 CME.  Results are shown for \replaced{progressively stricter sub-selections of data points that have distances above some fraction of the maximum radial extent of the simulated shock}{data sub-selections that are above progressively larger fractions of \textit{maxr}, the maximum radial distance of any point in that simulation time step}.  These outer data points naturally have a lower upstream plasma frequency, yielding a higher similarity to stencil derived from actual observed Wind/WAVES data. }}
\label{fig_frontSpectras}
\end{figure*}

\begin{figure*}[t]
\centering

\gridline{\fig{interpSyntheticFrontBit663.png}{0.47\textwidth}{Good geometrical Sub-selection, no radial sub-selection, 3x3 sector around $(6^{\circ}, -12^{\circ})$}
        \fig{interpSyntheticFrontBit690.png}{0.47\textwidth}{Bad geometrical sub-selection, no radial sub-selection, 3x3 sector around \replaced{$(-42^{\circ}, -3^{\circ})$}{$(-33^{\circ}, -9^{\circ})$}}
}

\caption{\textmd{Comparing two geometrical simulation features of the entropy derived shock showing how certain areas of the shock better fit the stencil of the type II burst determined from spacecraft observed data.  The scores for each of these geometrical sub-selections make up the similarity score histogram in Figure \ref{fig_heatmap},  \added{the upper left panel of which has a white circle and a red circle to indicate the good and bad regions (respectively) used to generate the synthetic spectra in this figure.}}}
\label{fig_goodbad}
\end{figure*}

\subsection{Geometrical Data Selections}

One may further refine these data extractions to test the similarity of synthetic spectra derived from different geometrically defined areas in the CME.  One can consider for each moment in time only keeping \replaced{the outermost portion of the selected model}{data points above some set fraction of the furthest out point with some radial distance \textit{maxr}} and discarding the rest.  This would in effect only leave the portion of the shock furthest out from the Sun, which would typically have lower upstream plasma densities, and therefore lower plasma frequencies.  One can see from Figure \ref{fig_frontSpectras} that \replaced{taking progressively more restrictive global radial sub-selections}{increasing the inner radial cutoff to 0.7*\textit{maxr} (upper right), 0.8*\textit{maxr} (lower left), or 0.9*\textit{maxr} (lower right)} in this manner can lead to \added{a} synthetic \replaced{spectra}{spectrum} with a higher similarity score, with more of its remaining data within the envelope of the stencil.  \deleted{Here, only data points furthest out from the Sun are counted, with a time dependent radial threshold of 70\%, 80\%, or 90\% of the maximum distance traveled by any point in the entropy enhanced feature at that time.}\added{  The fact that this is true for the entropy based sub-selection and not the density or magnetosonic mach $\#$ sub-selections is significant.  It signals that the outer edge of the entropy derived shock is a potential site of radio burst emission.  Here we take a moment to emphasize that this framework is a diagnostic tool, and is not expected to provide a guarantee of particle acceleration around simulated regions that provide high scoring synthetic spectra.  }

\begin{figure*}[t]
\centering

\gridline{\fig{NEWlongLatHeatmaps100-norm.png}{0.45\textwidth}{All (non-artifact) entropy enhanced data included}
        \fig{NEWlongLatHeatmaps70-norm.png}{0.45\textwidth}{Only data points with distance $>$ 0.7*\textit{maxr} included }
}

\gridline{\fig{NEWlongLatHeatmaps80-norm.png}{0.45\textwidth}{Only data points with distance $>$ 0.8*\textit{maxr} included}
        \fig{NEWlongLatHeatmaps90-norm.png}{0.45\textwidth}{Only data points with distance $>$ 0.9*\textit{maxr} included}
}

\caption{\textmd{Score distribution of synthetic spectra made from different geometrical sub-selections of entropy derived shock \added{over 120 minutes of simulation data}. Each cell in the histogram corresponds to a 3x3 degree subset of the simulation data whose synthetic \replaced{spectra}{spectrum} is created and similarity score calculated, using that score as the value on the histograms here, normalized to the maximum score of 107.  \added{The upper left panel has a white circle and a red circle to indicate the good and bad regions (respectively) of the shock used to make the synthetic spectra in Figure \ref{fig_goodbad}.}}}
\label{fig_heatmap}
\end{figure*}

Each of the regions across the shock surface can then be subdivided even further to better isolate potential locations of radio emission.  By looking at sub-selections based on solar longitude and latitude to create synthetic spectra, one can test hypotheses of stationary (with respect to the outward moving shock structure) burst production sites.  Synthetic spectra from two geometrical simulation features performed on the entropy enhanced data with no global radial cutoff can be seen in Figure \ref{fig_goodbad}.  The left panel shows the synthetic \replaced{spectra}{spectrum} from the portion of the shock within $-15^{\circ}$ - $-12^{\circ}$ longitude, $-9^{\circ}$ - $-6^{\circ}$ degrees latitude that clearly lies inside the defined stencil envelope.  The right panel shows the synthetic \replaced{spectra}{spectrum} from the portion of the shock within \replaced{$-45^{\circ}$ - $-42^{\circ}$ degrees longitude and $-3^{\circ}$ - $0^{\circ}$ degrees latitude}{$-33^{\circ}$ - $-30^{\circ}$ degrees longitude and $-9^{\circ}$ - $-6^{\circ}$ degrees latitude} that clearly lies mostly outside the defined stencil envelope.  This comparison shows how certain areas of the shock better fit the stencil of the type II burst determined from spacecraft observed data.

In addition to studying synthetic spectra from single sectors of the entropy based shock, looking at the geometrical distribution of scores of the event can be enlightening as well.  Figure \ref{fig_heatmap} shows a 2D spatial color contour graph where each cell corresponds to the similarity score for a synthetic \replaced{spectra}{spectrum} created for a 3x3 degree geometrical simulation feature of the simulation, normalized to the maximum score of 107\added{ over 120 minutes of simulation data}.  The upper left panel of Figure \ref{fig_heatmap} shows this similarity score distribution for all the non-artifact data in the entropy feature, showing that the area of the CME to the southwest of the Sun – Earth line around $6^{\circ}$ Longitude and $-12^{\circ}$ Latitude has a robust maximum.  \added{This panel has a white circle and a red circle to indicate the good and bad regions (respectively) of the shock used to make the synthetic spectra in Figure \ref{fig_goodbad}.}

The other panels are also 2D similarity score graphs, but with progressively stricter global radial sub-selections, where only data points furthest out from the Sun are counted, with a time dependent radial threshold of 70\%, 80\%, or 90\% of the maximum distance traveled by any point in the entropy enhanced feature at that time.  All of these similarity score distributions with various radial cutoffs share a global maximum around $(6^{\circ}, -12^{\circ})$.  One can see that the similarity score distributions across the shock for global radial cutoff $r > 0.7$\textit{maxr} maintains the maximum normalized score achieved with the no global radial cutoff subset of roughly 0.73, seen in the top 2 panels of Figure \ref{fig_heatmap}.  However, the $r > 0.7$\textit{maxr} radial cutoff also has a region in the northwestern portion of the shock where the similarity scores have improved relative to the no global cutoff subset.  This implies that the $r > 0.7$\textit{maxr} cutoff is getting rid of some secondary population of points slightly behind the main shock with slightly higher plasma frequencies, reducing the similarity score for those regions in the uncut subset.  This $r > 0.7$\textit{maxr} cutoff subset will be used for the remainder of the plots unless stated otherwise.  Also apparent in these panels is that the feature of the shock around $(-30^{\circ}, 0^{\circ})$ that has a constant normalized similarity score of around 0.3 across the different cutoffs, indicating that this area of the shock is at the leading edge.  These eastern and southwestern edges of the shock will be more closely investigated in the following sections.

\subsection{Derived Plasma Parameters}

One can also use lists of potential key parameters from \cite{Pulupa2010} that are known to be correlated with upstream Langmuir waves to guide selection of simulation features.  This work focuses on the upstream plasma data, starting with examining the \replaced{total upstream velocity}{upstream velocity magnitude} and plasma frequency across the surface of the entropy enhanced MHD feature.  Figure \ref{fig_varHeatmapFreq} shows the 2D spatial color contour graphs of upstream plasma frequency and upstream \replaced{total velocity}{velocity magnitude} across the \added{entire }surface of the shock at 20, 40, and 90 minutes after the start of the simulation\added{, displaying the mean values of the parameters for each longitude \& latitude bin}.  The plasma frequency was calculated with the upstream simulated electron density, and is used in creation of the synthetic radio spectra.  The \replaced{total upstream velocity}{upstream velocity magnitude} is shown as an example of a variable that is fairly well correlated with \textit{in situ} observations of Langmuir waves, ranking 9/25th of the plasma parameters tested in \citet{Pulupa2010}.  The general pattern of \replaced{total upstream velocity}{upstream velocity magnitude} across the shock front of having higher values on the southwestern side of the shock, smoothly transitioning to lower values on the eastern edge of the shock, is shared with the upstream magnetic field magnitude and upstream Alfv\'en speed, other plasma parameters that are fairly well correlated with \textit{in situ} observations of Langmuir waves.  \added{The 2D spatial color contour graphs for upstream magnetic field magnitude and upstream Alfv\'en speed are not pictured here to save space.}  

\citet{Pulupa2010} found that shocks with lower upstream plasma densities are more likely to exhibit Langmuir wave activity, indicating that the upstream solar wind structure is important in forming the shape of the shock and in any acceleration of particles nearby.  This is another reason that it has taken this long for simulations to get to a point where studies like this are possible; only recently have simulations, like the one done in \cite{Manchester2014}, accurately simulated the steady state solar wind structure of the inner heliosphere specific to the period preceding a simulated CME, before launching said CME and propagating it though the heliosphere.

Another plasma parameter classically thought to play an important role in particle acceleration at CMEs (and then Langmuir wave creation, followed by type II burst generation) is $\Theta_{Bn}$, the angle between the shock normal $\hat{n}$ and the upstream magnetic field vector $B$.  \citet{Pulupa2010} has shown $\Theta_{Bn}$ to be fairly well correlated with energetic electrons, with higher angles near $90^{\circ}$ having a higher level of association with electron beams near shocks.  Another correlated parameter is de Hoffmann-Teller frame velocity, $ \mathbf{V}_{HT}$, which was found to be the plasma parameter most correlated with energetic electrons by \cite{Pulupa2010}. $\mathbf{V}_{HT}$ is defined by the following equation.   






\begin{align}
    \mathbf{V}_{HT} = \frac{ \bf{\hat{n}} \times \left( \mathbf{V}_{u} \times \bf{B}_u \right)}{\bf{B}_u \cdot \bf{\hat{n}}}
\end{align}

\begin{figure*}[h]
\centering

\gridline{\fig{Entvar3Heatmaps20.png}{0.43\textwidth}{}
        \fig{Entvar9Heatmaps20.png}{0.43\textwidth}{}
}

\gridline{\fig{Entvar3Heatmaps40.png}{0.43\textwidth}{}
        \fig{Entvar9Heatmaps40.png}{0.43\textwidth}{}
}

\gridline{\fig{Entvar3Heatmaps90.png}{0.43\textwidth}{}
        \fig{Entvar9Heatmaps90.png}{0.43\textwidth}{}
}

\caption{\textmd{2D spatial color contour graphs of average upstream \replaced{total velocity}{velocity magnitude} and upstream plasma frequency over the front edge of the entropy derived shock.}}
\label{fig_varHeatmapFreq}
\end{figure*}

\begin{figure*}[h]
\centering

\gridline{\fig{Entvar16Heatmaps20.png}{0.43\textwidth}{}
        \fig{Entvar17Heatmaps20.png}{0.43\textwidth}{}
}

\gridline{\fig{Entvar16Heatmaps40.png}{0.43\textwidth}{}
        \fig{Entvar17Heatmaps40.png}{0.43\textwidth}{}
}

\gridline{\fig{Entvar16Heatmaps90.png}{0.43\textwidth}{}
        \fig{Entvar17Heatmaps90.png}{0.43\textwidth}{}
}

\caption{\textmd{2D spatial color contour graphs of average de Hoffmann-Teller frame velocity and $\Theta_{Bn}$ over the front edge of the entropy derived shock.}}
\label{fig_varHeatmapTheta}
\end{figure*}

$\mathbf{V}_{HT}$ is closely related to $\Theta_{Bn}$, with the denominator, $\bf{B}_u \cdot \bf{\hat{n}} = cos(\Theta_{Bn}) |\bf{B}|$, approaching zero as  $\Theta_{Bn}$ nears $90^{\circ}$, causing $\mathbf{V}_{HT}$ to become very large.  Here, $\bf{V}_u$ is the upstream velocity of the plasma in the shock rest frame.  An approximation for $\bf{V}_u$ is made by taking $\bf{V}_u = \bf{V}_{sim} - 1500\bf{\hat{n}}$, for simplicity assuming a constant shock speed of 1500 km/s in the direction of the shock normal.  Here, $\bf{V}_{sim}$ is the simulated upstream plasma velocity in the regular HEEQ frame\added{whose magnitude is shown in Figure \ref{fig_varHeatmapFreq}}.  Figure \ref{fig_varHeatmapTheta} shows the 2D spatial color contour graphs of $\Theta_{Bn}$ and $\mathbf{V}_{HT}$ across the surface of the shock at 20, 40, and 90 minutes after the start of the simulation.  There is more structure within these variable contour graphs than the graph of \replaced{$\mathbf{V}_sim$}{$\mathbf{V}_{sim}$} in Figure \ref{fig_varHeatmapFreq}, with higher values around the edges of the shock that match well with the structure in the similarity score contour graphs.  The area of maximum similarity around $(6^{\circ}, -12^{\circ})$ coincides with an area of consistently high $\mathbf{V}_{HT}$ over the 2 hours of simulation time.  The other feature of note is the area of highly elevated $\mathbf{V}_{HT}$ around the eastern edge of the shock at $(-30^{\circ}, 0^{\circ})$, with $\mathbf{V}_{HT}$ falling off after about an hour into the simulation.







One may use these correlated plasma parameters to provide additional restrictions on the MHD features and examine the resulting synthetic spectra to ascertain which thresholds produce the best match to the actual Wind/WAVES \replaced{spectra}{spectrum}.  This is done in the following subsection with $\mathbf{V}_{HT}$, as its structure across the entropy derived shock seems to have the best match to the features seen in the uncut similarity score distribution in Figure \ref{fig_heatmap}.


\begin{figure*}[t]
\centering
\gridline{\fig{interpSyntheticFront0-90-VHT2000.png}{0.5\textwidth}{r $>$ 0.9*\textit{maxr}, $\mathbf{V}_{HT} > 2000$ km/s}
        \fig{interpSyntheticFront0-70-VHT2000.png}{0.5\textwidth}{r $>$ 0.7*\textit{maxr}, $\mathbf{V}_{HT} > 2000$ km/s}
}

\caption{\textmd{Left: derived synthetic \replaced{spectra}{spectrum} for front edge of shock front with $\mathbf{V}_{HT} > 2000$ km/s, using the r $>$ 0.9*\textit{maxr} cutoff. Right: derived synthetic \replaced{spectra}{spectrum} for $\mathbf{V}_{HT} > 2000$ km/s using the r $>$ 0.7*\textit{maxr} cutoff.}}
\label{fig_varHeatmapThetaSpectra}
\end{figure*}

\begin{figure*}[b]
\centering
\gridline{\fig{interpSyntheticFrontBit810-90-VHT2000.png}{0.5\textwidth}{3x3 sector around  $(-30^{\circ}, 0^{\circ})$, r $>$ 0.9*\textit{maxr}, $\mathbf{V}_{HT} > 2000$ km/s}
        \fig{interpSyntheticFrontBit663-VHT2000.png}{0.5\textwidth}{3x3 sector around $(6^{\circ}, -12^{\circ})$, r $>$ 0.7*\textit{maxr}, $\mathbf{V}_{HT} > 2000$ km/s}
}

\caption{\textmd{Left: derived synthetic \replaced{spectra}{spectrum} for the 3x3 degree sector of the simulation around $(-30^{\circ}, 0^{\circ})$ using the r $>$ 0.9*\textit{maxr} cutoff, and a $\mathbf{V}_{HT} > 2000.$ km/s cutoff.  Right:  derived synthetic \replaced{spectra}{spectrum} for the 3x3 degree sector of the simulation around $(6^{\circ}, -12^{\circ})$ using the r $>$ 0.7*\textit{maxr} cutoff, and a $\mathbf{V}_{HT} > 2000.$ km/s cutoff.}}
\label{fig_varHeatmapThetaBit}
\end{figure*}

\subsection{Sub-selections based on both Geometry and Plasma Parameters}

Starting at the broadest scales and working down, Figure \ref{fig_varHeatmapThetaSpectra} shows synthetic spectra from a further reduction of the global radial cutoff subsets by imposing a minimum threshold of $\mathbf{V}_{HT}> 2000$ km/s across the entire shock.  This value was chosen as it seems to partition nicely the areas of elevated $\mathbf{V}_{HT}$, and corresponds to roughly 1.3 times the simplified 1500 km/s shock speed used to calculate the $\mathbf{V}_{HT}$ values.  \added{Since $\mathbf{V}_{HT}$ is inversely proportional to $cos(\Theta_{Bn})$, adopting a threshold of 1.3 times the shock speed means that $1/cos(\Theta_{Bn}) > 1.3 \implies \Theta_{Bn} > 40^{\circ}$, which will cut out most of the low $\Theta_{Bn}$ data, but still be inclusive of elevated regions of $\Theta_{Bn}$, not just including the very highest.  \textit{In situ} shock analysis in \citet{Pulupa2010} Figure 3 shows that the cumulative distribution function for shocks that exhibit intense upstream Langmuir wave activity only really starts going up past 0.1 after $\Theta_{Bn} = 40^{\circ}$.   }These synthetic spectra show a broad overlap with the stencils, suggesting that the process creating the type II burst may be non-localized, with multiple points all across the shock passing some physical thresholds such as $\mathbf{V}_{HT} > 2000$ km/s to start the burst generation process.  However, analyzing the similarity distribution as a function of geometry may reveal that the emission is concentrated to distinct regions on the shock.  The following figures analyze these trends in the distribution of similarity scores in synthetic spectra that take both geometry and a threshold of $\mathbf{V}_{HT}> 2000$ km/s into account.

One may zoom in on that area of the shock that does best in stationary burst hypotheses, and see if it holds up with under various sub-selections based on thresholds of plasma parameters.  This is done with a cut of  $\mathbf{V}_{HT} > 2000$ km/s in Figure \ref{fig_varHeatmapThetaBit} where on the left is the synthetic \replaced{spectra}{spectrum} for the 3x3 degree sector of the simulation around $(-30^{\circ}, 0^{\circ})$ that matches the area of high $\mathbf{V}_{HT}$ in the first hour of the simulation, and on the right is the synthetic \replaced{spectra}{spectrum} for the 3x3 degree sector of the simulation around $(6^{\circ}, -12^{\circ})$ with a high degree of similarity to the Wind data.   On the left, one can see that this eastern edge of the shock matches the stencil quite well before tapering off after about an hour.  On the right, one can see that the similarity score around the $(6^{\circ}, -12^{\circ})$ edge is nearly unchanged from the left panel of Figure \ref{fig_goodbad}, lowering from 82 to 67, only a 20\% dip, indicating that a population of points within this geometrical region of the shock with high $\mathbf{V}_{HT}$ could be where at least part of the burst generation occurred.   

\begin{figure*}[t]
\centering
\gridline{\fig{NEWlongLatHeatmaps90-norm-VHT2000-1hour.png}{0.5\textwidth}{1 hour normalized, r $>$ 0.9*\textit{maxr}, $\mathbf{V}_{HT} > 2000$ km/s}
        \fig{NEWlongLatHeatmaps70-norm-VHT2000-1hour.png}{0.5\textwidth}{1 hour normalized, r $>$ 0.7*\textit{maxr}, $\mathbf{V}_{HT} > 2000$ km/s}
}
\gridline{\fig{NEWlongLatHeatmaps90-norm-VHT2000.png}{0.5\textwidth}{2 hour normalized, r $>$ 0.9*\textit{maxr}, $\mathbf{V}_{HT} > 2000$ km/s}
            \fig{NEWlongLatHeatmaps70-norm-VHT2000.png}{0.5\textwidth}{2 hour normalized, r $>$ 0.7*\textit{maxr}, $\mathbf{V}_{HT} > 2000$ km/s}
}

\caption{\textmd{Top: similarity score contour graphs calculated using only the first hour of data.  Left: $r > 0.9$\textit{maxr}, $\mathbf{V}_{HT} > 2000.$  Right: $r > 0.7$\textit{maxr}, $\mathbf{V}_{HT} > 2000.$ Bottom: similarity score contour graphs calculated using all 2 hours of data. Left: similarity score contour graph for $r > 0.9$\textit{maxr}, $\mathbf{V}_{HT} > 2000.$ km/s  Right:  similarity score contour graph for $r > 0.7$\textit{maxr}, $\mathbf{V}_{HT} > 2000.$ km/s }}
\label{fig_varHeatmapVHT} 
\end{figure*}

One can also do a geometrical analysis of similarity scores while also including the variable threshold sub-selection $\mathbf{V}_{HT} > 2000$ km/s, the results of which are shown in Figure \ref{fig_varHeatmapVHT}.  The top 2 panels show similarity score 2D spatial color contour graphs calculated using only the first hour of simulation data, and are normalized as such.  The top left panel shows the distribution using the cutoffs r $>$ 0.9*\textit{maxr} and $\mathbf{V}_{HT} > 2000$ km/s, while the top right shows the distribution using the cutoffs r $>$ 0.7*\textit{maxr} and $\mathbf{V}_{HT} > 2000$ km/s.  The bottom 2 panels show similarity score contour graphs calculated using the same respective cutoffs, but using all 2 hours of simulation data, and are normalized as such.  So when comparing scores of the top row panels and bottom row panels, one must take roughly half of the upper scores to compare them on the same scale with the bottom scores.

From these contour graphs one may surmise a few things.  First, the eastern edge of the shock around $(-30^{\circ}, 0^{\circ})$ which coincides with a large swath of high $\mathbf{V}_{HT}$, is at the leading front of the shock for the first hour of the event, and matches the observed burst envelope quite closely in this period (see Figure \ref{fig_varHeatmapThetaBit} a).  By looking at the two left panels, one can see that after 1 hour, the eastern edge of the shock does not gain any additional points to its similarity score, indicating that after that an hour, not much type II emission is originating from there.  This can also be clearly seen in the synthetic \replaced{spectra}{spectrum} for this area of the shock in the left panel in Figure \ref{fig_varHeatmapThetaBit}.  This area of the shock also has high $\Theta_{Bn}$ and $\mathbf{V}_{HT}$ values for the first hour of the simulation, as seen in the variable 2D spatial color contour graphs in Figure \ref{fig_varHeatmapTheta} a.  This eastern edge of the shock can also been seen as a feature in the similarity score histograms, being most prominent in the r $>$ 0.9\textit{maxr} sub-selections.  One can also see an obvious maximum around $(6^{\circ}, -12^{\circ})$ on the bottom two panels, and a feature at that area in the top right panel, coinciding with the previous similarity score contour graphs around the southwestern edge of the CME shock.  This corresponds to areas where there is a strip of consistently high $\Theta_{Bn}$ and $\mathbf{V}_{HT}$ on the southern and western edges of the shock.

A picture emerges where the key emitting parts of the shock are likely those with high $\mathbf{V}_{HT}$, with the eastern edge of the shock in the fist 40 minutes or so dominating the emission, then falling off, not being shocked strongly enough past a certain point, or some change in upstream magnetic field decreasing the local $\mathbf{V}_{HT}$.  The western and southern edges of the shock are also seen to have points with consistently high $\mathbf{V}_{HT}$, with a maximum similarity centered around  $(6^{\circ}, -12^{\circ})$, roughly the southwestern corner of the main shock.  This high $\mathbf{V}_{HT}$ signature along much of the western edge of the shock persists through all but the beginning of the 2 hours of simulation time, achieving a high similarity score across a wide area in longitude and latitude before the  $\mathbf{V}_{HT}$ based sub-selection.  After the $\mathbf{V}_{HT}$ based sub-selection, the maximum around  $(6^{\circ}, -12^{\circ})$ is more pronounced, but the area around it remains relatively high, indicating that there may be some level of type II emission from much of the shock or shock edge where there exist areas of high $\mathbf{V}_{HT}$.


\section{Possible Sources of Error}


MHD simulations have many limitations and simplifying assumptions and this one is no exception.  A significant source of error arises in how the flux rope is inserted into the simulation to recreate a solar eruption and coronal mass ejection.  The CME is initiated by inserting a Gibson-Low magnetic flux rope \citep{Gibson_1998} into the corona that is line-tied to the bottom boundary at the location of active region 10759 at 16:03 UT.  The specific deformation of the flux rope  (a mathematical stretching to draw the flux rope towards its origin, while preserving its angular size) leaves the flux rope in a state of force imbalance causing an instantaneous  eruption.  This treatment of the flux rope fails to capture the energy buildup and early acceleration phase of the eruption.  There is also the matter of magnetic reconnection, which is done in the model through numerical diffusion, which is likely leading to reconnection rates faster than what would actually happen.  More recent simulations by \cite{Torok2018} capture the liftoff phase.  Nonetheless, the important aspects that the simulation needs to capture for this study are the speed profile of the CME and the ``upstream'' mass density within 20 \replaced{Rs}{solar radii} to calculate the plasma frequency.  \replaced{Seen in Figure \ref{fig_shockSpeed}}{As discussed in Section 6}, the initial speed of the simulated CME is just above 2500 km/s, close to the estimate of \citet{Reiner2007}, which used the LASCO observations as an initial plane of sky estimate of the CME speed, before fitting the kinematic profile with the radio data.  In some sense, the best test of the simulation's accuracy of the density structure close to the Sun is how well the synthetic \replaced{spectra}{spectrum} generated by the shock, or whatever the emitting structure is, match the real data.  As is seen in the previous sections, the synthetic spectra from simulations of the entropy derived shock largely match the observed radio data, lending confidence to the simulation's accuracy near the Sun.         


There is also an argument to be made that it is not appropriate to compare a synthetic \replaced{spectra}{spectrum}, which is really a column normalized 2D histogram of the plasma frequency of points in a selected feature over the simulation domain, to an actual radio \replaced{spectra}{spectrum} which has physical units of spectral flux density.  This would be a legitimate argument if this framework were trying to deduce the exact physical mechanism behind particle acceleration and type II burst generation and propagation all at once in this single work.  This is not the case.  This work is trying to provide physical thresholds of plasma parameters that when surpassed, seem to create spectra similar to entropy enhanced  those actually observed.  If these thresholds are consistent across simulations and observations of multiple events, that will be a significant takeaway that can inform forecasting models of Earthbound energetic particles.  It would also provide heuristic checks that can be performed on fully fleshed out theories of particle acceleration and type II burst generation, potentially ruling out entire classes of theories.  

There is also the issue that the model being used includes an adaptive mesh refinement (AMR) scheme.  This implies that each point being counted may be of a different size.  In this case, the blocks are refined in a spherical wedge centered on the trajectory of the CME, with blocks closer to the Sun being more highly refined. We argue that this fact is not an issue for the purpose of type II burst template matching.  Many of the points identified in an initial phase of the eruption come from areas of high entropy that may be more highly refined than regions of the weaker shock.  In general, the relatively small frequency bandwidth observed from type II bursts suggest that the emission source is localized or occurs from a region of uniform density and thus uniform radial distance from the Sun.  Therefore it is likely that the model emission sources that are successful in recreating the observed radio spectrum come from localized areas of similar density and plasma frequency that are likely to be of the same level of refinement.  Thus, for the proposed purposes of identifying useful physical thresholds or geometrical criteria that correspond to type II burst generation, apples to oranges comparison between synthetic spectra and spacecraft observed spectra is not a fatal flaw.     

\section{Discussion and Future Work}

In this work, the simulated structures of the 2005 May 13 radio-loud ICME were compared with \added{the} Wind/WAVES recorded radio \replaced{spectra}{spectrum} to identify likely regions where the type II burst was generated.  This work attempts to solve the inverse problem of the location of the burst generation site by analyzing the plasma parameters in detail from an MHD simulation of the real event.  Different physical structures (shock and current sheets) of the MHD simulations were extracted to create time-varying synthetic radio spectra and were compared to actual Wind/WAVES radio data to determine which hypothesized source region best matches the spacecraft observed data.  

An event specific ``stencil'' was developed and used to assign a similarity score to the synthetic spectra based on how much of its data is aligned with the observed Wind/WAVES data.  The stencil is an estimated spectral time profile employing a Gaussian intensity profile fitted over the identified type II fundamental emission.  This fundamental emission was identified using a kinematic profile of the CME made by \cite{Reiner2007}.  The larger the portion of data from a given MHD feature whose upstream plasma frequency lies within this stencil, the higher similarity score it will receive.  This scoring stencil is described in depth in section 4.

This methodology bypasses the need for simulating the exact physics that are creating the bursts, and outputs what areas of the CME are most likely to be the site of particle acceleration.  These locations were then inspected more closely with the additional sub-selections to the simulation data to see which plasma parameters are important for type II burst generation to occur.  This methodology combines remotely sensed type II radio data alongside virtual \textit{in situ} measurements through simulation data, enabling a detailed look at the plasma parameters at the best fit for the site of emission.  

This technique was employed with several different model extractions to test the feasibility of different classes of particle acceleration mechanisms.  The \replaced{three}{four} main broad features used here were points with at least a 3.5 MK temperature used to identify the current sheet, points that were shocked to a factor of 3.5 or more in their density compared to before the shock wave, \deleted{and }points that had their entropy enhanced by a factor of 4 or more compared to before the event\added{, and points that have a Fast Magnetosonic Mach $\# \geq 3$}.  Each of these data sources had 2 hours of simulation data, with a time cadence of 2 minutes in the beginning of the run, transitioning to every 5 minutes later in the run for a total of 35 snapshots over 120 minutes.  \added{The one exception to this simulation sampling cadence is the selection for Fast Magnetosonic Mach $\# \geq 3$ sub-selection, which due to a lack of resources only combines the six times of [10, 20, 30, 40, 60, 90] minutes after the start of the simulation.  }Synthetic spectra showing the upstream plasma frequency of these \replaced{three}{four} broad MHD features alongside the spacecraft observed radio data reveal that of the \replaced{three}{four} broad synthetic spectra, the one from the entropy enhanced feature best matches the Wind/WAVES observations both in spectral width and overall shape.  This similarity is reflected in the computed similarity scores of each of these synthetic spectra, with the entropy feature having the highest similarity score.

With these broad MHD features in hand, finer sub-selections were also applied based on plasma parameters thought to be important for SEP acceleration, starting with sub-selections based on the geometry of the CME.  The framework created here explores the optimization problem with a stationary burst assumption, one where the burst location has a constant longitude and latitude and continues outward with the outer edge of the entropy derived shock.  An analysis of the similarity scores for different geometrical simulation features from the entropy enhanced data over the 2 hours of simulation time indicates that the area of the shock near $(6^{\circ}, -12^{\circ})$ has the greatest overall similarity with the main band of fundamental emission in the stencil identified by the kinematic solution of the CME.

In addition to investigating the similarity of synthetic spectra generated from different geometrical regions across the entropy derived shock, further constraints on the data are considered, namely a threshold based on de Hoffmann Teller speed ($\mathbf{V}_{HT}$).  This plasma parameter is chosen both for its high correlation with \textit{in situ} observations of Langmuir waves \citep{Pulupa2010}, and also for the fact that the distribution of this parameter across the surface of the shock generally matches the elevated features seen in the similarity score distributions around the eastern and southwestern edges of the shock, more so than other parameters like \replaced{total upstream velocity}{upstream velocity magnitude}.    
The spatial distribution of the similarity scores after an additional constraint of $\mathbf{V}_{HT} > 2000 $km/s was also examined, where it is found that the key areas of high similarity are unchanged, the matching peaks lending confidence to the suitability of using enhancements in de Hoffmann-Teller speed as a predicting factor of particle acceleration at CME driven shocks.  Overall, the areas with the highest similarity in their generated synthetic spectra coincide with areas of high $\Theta_{Bn}$ and $\mathbf{V}_{HT}$ on the southern and western edges of the shock, as well as the eastern edge for the first hour.  This shared structure was not present in other plots of plasma parameters across the entropy derived shock structure.  The fact that the structure of de Hoffmann-Teller speed has the best match to the similarity structure of synthetic spectra across the shock is not particularly surprising, since \citet{Pulupa2010} found that this same parameter was most highly correlated with \textit{in situ} observations of Langmuir waves at 1 AU.

In the future, additional studies utilizing this analysis framework across other historical events can be used to find common physical thresholds that create synthetic spectra that match the observed \replaced{spectra}{spectrum}.  If these thresholds are consistent across simulations and observations of multiple events, that will be a significant takeaway that can inform forecasting models of Earthbound energetic particles.  It would also provide heuristic checks that can be performed on fully fleshed out theories of particle acceleration and type II burst generation, potentially ruling out entire classes of theories.  

This analysis framework would also become more powerful with the addition of multiple spacecraft for imaging or localization at these low radio frequencies.  Single antenna observations can measure the total flux density of solar radio bursts, but cannot pinpoint where the emission is coming from.  Previous efforts such as \citet{Krupar2014} have been made in localizing low frequency emission from type III radio bursts using the two STEREO spacecraft, applying goniopolarimetric analysis techniques to deduce a wave vector direction and apparent source size at each spacecraft.  Combining these measurements from two or more spacecraft allows approximate triangulation of the source.  However, these direction finding techniques are not a substitute for true coherent imaging that is enabled at low frequencies by interferometry, as imaging yields information of the sky brightness pattern up to the resolution of the array.  The imaging approach will be taken by the Sun Radio Interferometer Space Experiment (SunRISE) \citep{alibay2018, Hegedus2019-1}, which will utilize 6 spacecraft equipped with dual-polarization low frequency radio antennas to form a radio interferometer in space.  If reliable localizations of the source of the type II emission are available from SunRISE or other means, this MHD analysis framework allows probing of the plasma parameters at the observed location of the emission, enabling in depth studies of the physics of acceleration.  At least one future paper will be dedicated to the combination of this MHD analysis framework with a simulated SunRISE data analysis pipeline to prove the feasibility of such an array, and hopefully more studies with actual data after SunRISE launches in 2023.  The combination of observations of multiple type II events with SunRISE alongside simulated MHD recreations utilizing this framework will hopefully lead to a satisfying answer to the question of how CMEs generate type II bursts and accelerate particles. 

\section*{Acknowledgments}

A. Hegedus, W. Manchester, and J. Kasper are supported by NASA funds for SunRISE 80GSFC18C0009.  A. Hegedus and J. Kasper are also supported and by the NASA Solar System Exploration Research Virtual Institute cooperative agreement number 80ARC017M0006, as part of the Network for Exploration and Space Science (NESS) team out of University of Colorado Boulder 1555631. J. Kasper is additionally supported by the NASA Wind fund, 80NSSC18K0986.  W. Manchester is also supported by NSF PRE-EVENTS grant No. 1663800 and by NASA grants NNX16AL12G, and 80NSSC17K0686.  The CME simulation was performed with the Pleiades system provided by NASA’s High-End Computing Program under award SMD-11-2364, and the NCAR Yellowstone supercomputer.  This research made use of NumPy \citep{harris2020array}.  This research made use of matplotlib, a Python library for publication quality graphics \citep{Hunter:2007}.  The authors would like to thank the many individuals and institutions who contributed to making Wind/WAVES possible. The authors would also like to thank the reviewers for their helpful suggestions.


\bibliography{sample63}{}
\bibliographystyle{aasjournal}



\end{document}